\documentclass[sn-apa]{sn-jnl}

\usepackage{palatino} 

\usepackage[T1]{fontenc}

\usepackage{graphicx}%
\usepackage{multirow}%
\usepackage{amsmath,amssymb,amsfonts}%
\usepackage{amsthm}%
\usepackage{mathrsfs}%
\usepackage[title]{appendix}%
\usepackage{xcolor}%
\usepackage{textcomp}%
\usepackage{manyfoot}%
\usepackage{booktabs}%
\usepackage{algorithm}%
\usepackage{algorithmicx}%
\usepackage{algpseudocode}%
\usepackage{listings}%

\theoremstyle{thmstyleone}%
\theoremstyle{thmstyletwo}%

\theoremstyle{thmstylethree}%

\begin{document}

\title[Article Title]{Non-invasive Assessment of Pancreatic Duct Hypertension Using Computational Flow Modeling}


\author[1]{\fnm{Haobo} \sur{Zhao}}\email{hzhao67@jh.edu}
\author[1]{\fnm{Jung-Hee} \sur{Seo}}\email{jhseo@jhu.edu}
\author[2]{\fnm{Venkata} \sur{Akshintala}}\email{vakshin1@jhmi.edu}
\author[2]{\fnm{Surya} \sur{Evani}}\email{sevani1@jh.edu}
\author*[1]{\fnm{Rajat} \sur{Mittal}}\email{mittal@jhu.edu}

\affil*[1]{\orgdiv{Mechanical Engineering}, \orgname{Johns Hopkins University}, \orgaddress{\street{3400 N. Charles Street}, \city{Baltimore}, \postcode{21218}, \state{MD}, \country{USA}}}
\affil[2]{\orgdiv{Department of Gastroenterology}, \orgname{Johns Hopkins Hospital}, \orgaddress{\street{600 N Wolfe St.}, \city{Baltimore}, \postcode{21287}, \state{MD}, \country{USA}}}

\abstract{Chronic pancreatitis (CP) is a progressive inflammatory disease frequently associated with severe, treatment-resistant abdominal pain, which is hypothesized to result from pancreatic ductal hypertension (PDH) secondary to ductal strictures or obstructions. However, direct measurement of pancreatic duct pressure (PDP) remains technically demanding and invasive, thereby significantly limiting its routine clinical application. Here, we propose and validate a novel, non-invasive approach for estimating PDP. The method integrates patient-specific magnetic resonance cholangiopancreatography (MRCP) imaging with computational fluid dynamics (CFD) modeling. Three-dimensional ductal models reconstructed from MRCP data enabled simulation of intraductal pressure distributions with high anatomical fidelity. The simulated pressure gradients showed strong correlation with in vivo measurements obtained via endoscopic retrograde cholangiopancreatography (ERCP), as well as with clinical outcomes such as pain relief following ductal decompression. To improve clinical usability, we developed a quasi-one-dimensional analytical model that accurately predicted PDP from ductal geometry and flow parameters, showing strong concordance with CFD results. These findings establish the feasibility and clinical relevance of MRCP-based PDP estimation, and underscore its potential as a non-invasive diagnostic tool for detecting PDH and informing therapeutic decisions in patients with CP.}




\maketitle


\newpage

\section{Introduction}

\begin{figure}[H]
\centering
\includegraphics[width=0.8\textwidth]{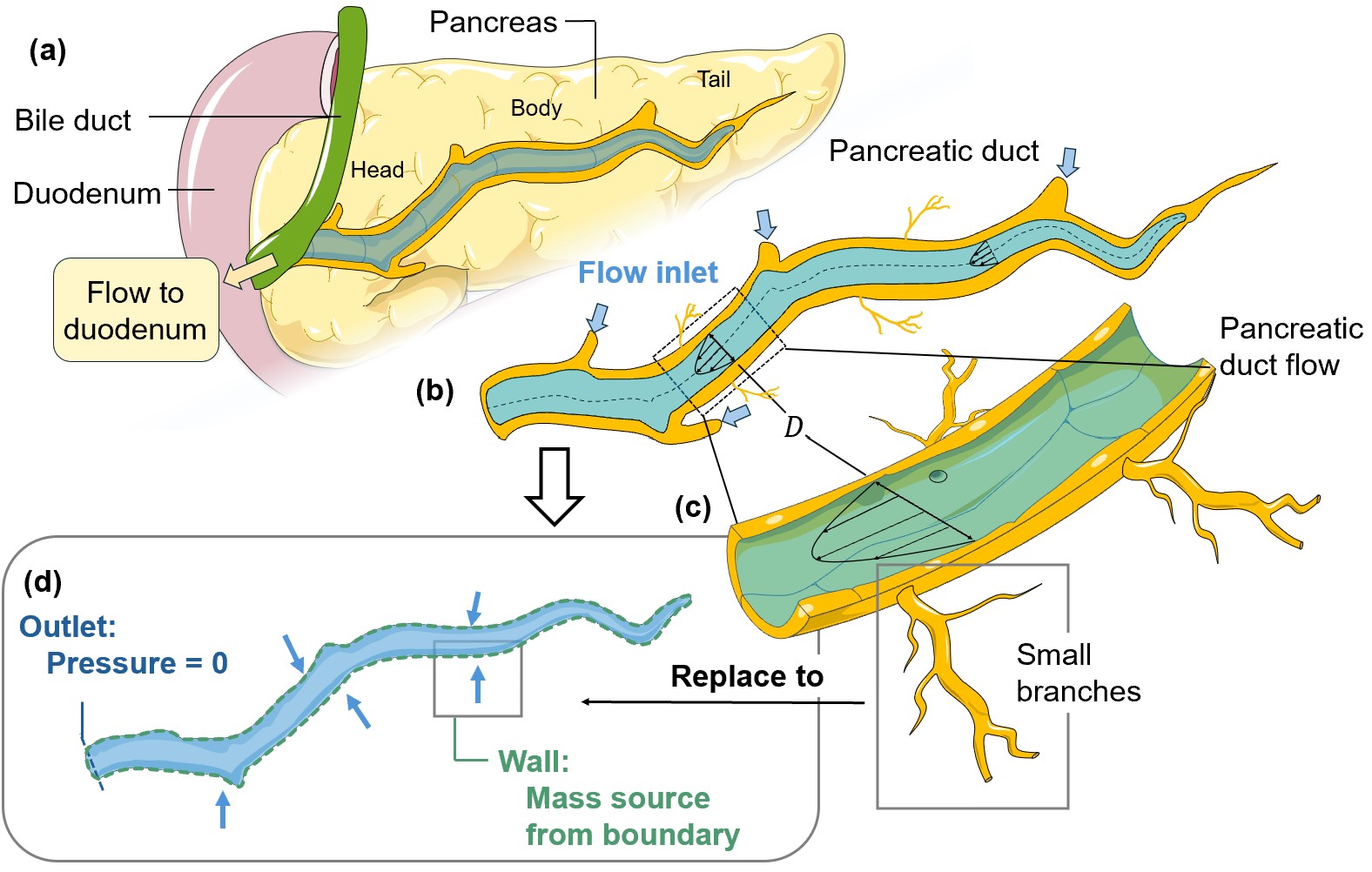}
\caption{%
Anatomical configuration and computational modeling of the pancreatic duct system. 
(a) Overview of the anatomical relationship among the pancreas, bile duct, and duodenum, highlighting the path of pancreatic juice flow toward the duodenum. 
(b) Schematic of overall pancreatic duct (PD) flow from tail to head, with secretions entering from small lateral branches. 
(c) Close-up of the PD segment with representative duct diameter ($D$) and typical anatomical features, including tributary branches. 
(d) Boundary conditions applied in computational fluid dynamics (CFD) simulations: flow inlet along the duct wall represents distributed secretion sources, and a zero-gauge pressure Dirichlet condition is imposed at the outlet near the duodenum. 
Typical PD length ($L$): 0.095–0.25 m; diameter ($D$): 0.002–0.005 m.
}
\label{figanatomy_BC}
\end{figure}

The pancreas is a retroperitoneal glandular organ that performs essential exocrine and endocrine functions. It produces digestive enzymes via its acinar cells and delivers them into the duodenum through the pancreatic duct (PD), while also regulating glucose metabolism through the secretion of insulin and glucagon. Anatomically, the PD extends longitudinally from the tail to the head of the pancreas, with a variable diameter that typically ranges from 1–3.5 mm in individuals under 50 years of age, and from 2–5 mm in older adults (70–79 years). The overall length of the PD ranges from 9.5 to 25 cm \citep{singh2023pancreatic}. The relative anatomical positioning is illustrated in Figure \ref{figanatomy_BC}.

\noindent
\setlength{\fboxrule}{0.5pt}
\setlength{\fboxsep}{10pt}
\fbox{%
\begin{minipage}{0.9\textwidth}
\textbf{Abbreviations:}\\[6pt]
\begin{small}
\begin{tabular}{@{}ll@{}}
CP   & Chronic pancreatitis \\
ERCP & Endoscopic retrograde cholangiopancreatography \\
MRCP & Magnetic Resonance Cholangiopancreatography \\
PD   & Pancreatic duct \\
PDP  & Pancreatic duct pressure \\
PDH  & Pancreatic ductal hypertension \\
\end{tabular}
\end{small}
\end{minipage}
}

\bigskip

Chronic pancreatitis (CP) is a progressive fibroinflammatory disease of the pancreas characterized by irreversible morphological changes and functional impairment. It frequently manifests as persistent abdominal pain, exocrine insufficiency, and endocrine dysfunction. The global incidence of CP ranges from 5 to 14 per 100,000 individuals annually, while prevalence estimates vary from 30 to 50 per 100,000, with some regions reporting rates exceeding 120 per 100,000 due to extended survival and under-diagnosis \citep{kleeff2017chronic}. In the United States, data from insured populations between 2001 and 2013 indicate a comparable incidence (4–5 per 100,000) and prevalence (25.4–98.7 per 100,000) \citep{beyer2020chronic}.

Despite declining rates of alcohol consumption and smoking — two principal risk factors — the incidence of CP has continued to rise. This epidemiological trend underscores the need for earlier diagnosis and more precise stratification of disease mechanisms.

A hallmark clinical feature of CP is chronic, often debilitating, abdominal pain. This pain is hypothesized to originate primarily from two distinct mechanisms: (1) pancreatic ductal hypertension (PDH) due to obstruction by strictures or stones, which raises intraductal pressure; and (2) pancreatic neuropathy, involving inflammation-induced perineural invasion and nerve remodeling \citep{ceyhan2009pancreatic}. Importantly, only PDH-related pain may respond to decompressive interventions such as endoscopic stenting or surgical ductal drainage, while neuropathic pain tends to be refractory. Consequently, objective and reliable identification of PDH is essential for appropriate patient selection prior to interventional procedures.
At present, direct in-vivo measurement of pancreatic duct pressure (PDP) relies on invasive techniques including ERCP-based perfusion manometry \citep{madsen1982intraductal,sato1986role}, microtransducer catheters \citep{vondrasek1974semiconductors}, and fistula-based pressure readings \citep{duval1958effect}. Although technically feasible, these methods are associated with high complication risks—including pancreatitis and ductal injury—and are unsuitable for routine clinical use or screening \citep{singh2023pancreatic}.
Given these limitations, there remains a critical unmet clinical need for a non-invasive, image-based method capable of estimating intraductal pressure and functionally characterizing pancreatic ductal hypertension (PDH).

Recent advances in computational biomechanics and image-derived flow modeling have demonstrated the feasibility of simulating hemodynamic parameters from cross-sectional imaging data. Patient-specific computational fluid dynamics (CFD) has been successfully applied in cardiovascular \citep{taylor2013computational,seo2014effect,mittal2016computational}, hepatic \citep{li2024impact}, and pulmonary systems \citep{atzeni2021computational}, enabling virtual diagnostics and interventional planning. Parallel developments in MRCP acquisition and processing—such as secretin-enhanced MRCP~\citep{bali2005quantification}, deep-learning-based pancreas segmentation~\citep{zhang2020deep}, and automatic 3D reconstruction~\citep{lechelek2022hybrid}—further support functional analysis of the pancreatic duct. In spite of these advances, non-invasive estimation of pancreatic ductal pressure has not yet been integrated into routine clinical workflows. Existing approaches primarily emphasize anatomical or morphological features, without direct validation of simulated pressure against in vivo measurements. To date, no study has quantitatively validated image-based CFD predictions of pancreatic ductal pressure using direct physiological measurements.

In this study, we introduce and validate a novel framework that integrates MRCP imaging with computational fluid dynamics (CFD) modeling to non-invasively quantify PDH. Using MRCP-derived 3D duct reconstructions, we simulate intraductal flow and compute pressure gradients, which are validated against ERCP-based pressure measurements. To enable broader clinical application, we also develop a quasi-one-dimensional analytical model to rapidly estimate pressure from geometry alone. Finally, we evaluate the association between simulated pressure and pain relief following ductal decompression, highlighting the potential clinical utility of this method in patient stratification and treatment planning.

\section{Method}

\subsection{3D Model Reconstruction}
We have developed a noninvasive approach to evaluate pancreatic ductal hypertension (PDH) by integrating patient-specific Magnetic Resonance Cholangiopancreatography (MRCP) data with computational fluid dynamics (CFD). High-resolution MRCP images were segmented and reconstructed into three-dimensional (3D) models of the pancreatic duct using the software, MIMICS and 3-MATIC (Materialise NV, Belgium), following standard radiological reference markings. 
Variability in MRI quality and patient anatomy necessitated post-processing to produce simulation-ready geometries. Cavity filling and surface smoothing algorithms were applied to eliminate topological artifacts and suppress noise. Small branches were manually excluded, and the terminal duct outlet was closed and smoothed to facilitate CFD meshing. The final geometries preserved anatomical accuracy while ensuring numerical convergence.


\begin{figure}[H]
\centering
\includegraphics[width=\textwidth]{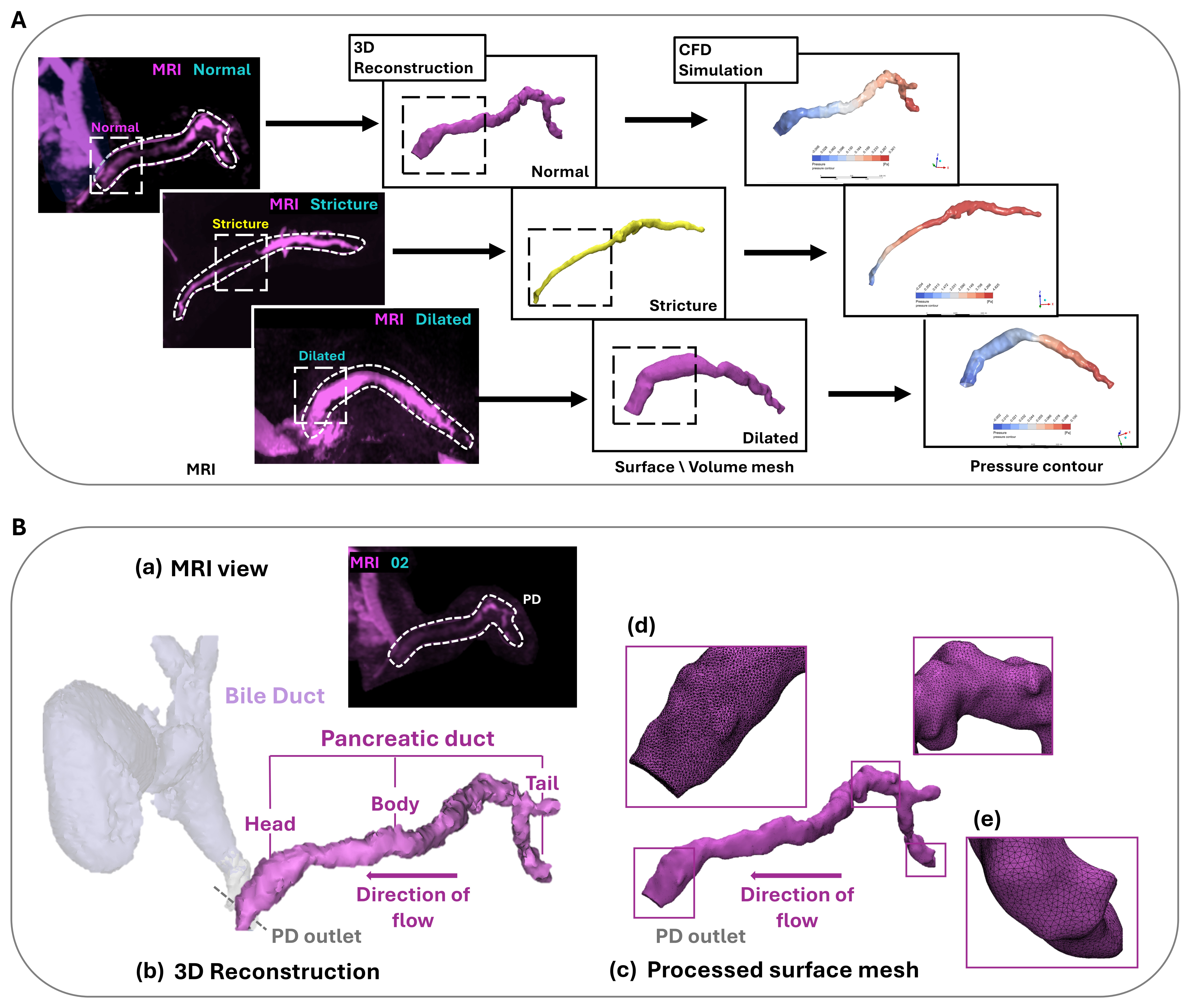
}
\caption{
\textbf{A}: Workflow for pancreatic duct types: normal PD, PD with strictures, and dilated PD (without stricture segment).
The diagram illustrates the identification of ductal abnormalities using MRI, followed by 3D reconstruction, computational meshing, and fluid dynamics simulation. 
Quantitative measurements are used to assess alterations in flow and the potential development of pancreatic duct hypertension.  
\textbf{B}: (a) MRI view; (b) 3D reconstruction; (c) and (d) processed surface mesh of the pancreatic duct model. 
This panel shows a representative example of the reconstructed duct geometry and the corresponding mesh following refinement, with anatomical labeling for further CFD processing.
}
\label{fig:reconstruction}
\end{figure}

Following the initial 3D reconstructions, 
we identified distinct anatomical scenarios of pancreatic duct, and the model development was focused to capture the following three kind of representative scenarios: a normal duct, a duct with a stricture, and a dilated duct (without stricture segment). 
As shown in Figure~\ref{fig:reconstruction}A, the complete workflow illustrates the process from MRCP-based identification of ductal abnormalities through to model reconstruction, mesh generation, and fluid dynamics simulation. This procedure enables quantitative assessment of pancreatic ductal hypertension (PDH) across various pathologies.
Figure~\ref{fig:reconstruction}B presents a representative example of the reconstructed duct geometry and its post-processed surface mesh. Refinement steps, including remeshing and smoothing, were applied to ensure numerical robustness while preserving anatomical accuracy. These high-fidelity models served as the basis for subsequent CFD analysis.

\subsection{Modeling Pancreatic Duct Flow}

The reconstructed models were assumed to represent fluid-filled, steady-flow systems and were meshed to permit volumetric flow simulations. Mesh generation was guided by anatomical fidelity and numerical stability requirements. All ducts were modeled under physiological conditions, assuming Newtonian, incompressible flow with no-slip boundary conditions applied to the duct walls. 
Pancreatic juice was modeled as a Newtonian fluid with density and viscosity parameters equivalent to water, consistent with \cite{pandol2010water}, which describes pancreatic juice as a clear, isotonic fluid primarily composed of water and electrolytes, indicating comparable physical properties in physiological and even mild CP states. The Reynolds number in the pancreatic duct is typically on the order of 10 \citep{tajikawa2023investigation}, justifying the assumption of laminar, steady-state flow.
We computed the pressure drop ($\Delta$P) along the duct as a surrogate for PDH. This approach has been inspired by FFR-CT \citep{taylor2013computational}, the computational framework that predicts the pressure drop across coronary artery lesions using CFD simulations based on cardiac computed tomography, and which has revolutionized the diagnosis and management of coronary artery disease.

To simulate the flow of pancreatic juice into and through the PD, appropriate physiological boundary conditions were imposed, as illustrated in Figure~\ref{figanatomy_BC}. Pancreatic secretions enter through small side branches distributed along the duct wall and exit via the common channel at the duct–bile duct junction, flowing into the duodenum (Figure~\ref{figanatomy_BC} (b),(c)).
Due to the difficulty in resolving thin ductal branches from MRCP, physiological flow was modeled using distributed mass sources along the duct walls. 
 In the CFD model, this was implemented as a uniformly distributed mass source along the duct wall which was 
treated as rigid and porous. 
 CFD simulations were performed on reconstructed ducts under varying mass influx conditions, simulating physiological secretion across the duct wall. 
Mass fluxes of 0.01, 0.05, 0.1, and 0.2 g/s were imposed, corresponding to physiologic pancreatic juice secretion rates ranging from 0.1 to 5 mL/min, as reported by \cite{pandol2010water}. 
A Dirichlet boundary condition with zero gauge pressure (0 Pa) was applied at the ductal outlet to mimic physiological drainage (Figure~\ref{figanatomy_BC} (d)). This also establishes a reference point for pressure drop measurement.

The governing equations for the incompressible, laminar, steady-state flow of pancreatic juice within the duct are the Navier–Stokes equations:
\begin{equation}
    \left\{
    \begin{aligned}
    & \nabla\cdot\mathbf{u}=0 \\
    & \rho\left[\frac{\partial\mathbf{u}}{\partial t}+\nabla\cdot(\mathbf{u}\mathbf{u})\right]
=-\nabla p+\mu\nabla^2\mathbf{u}
    \end{aligned}
    \right.
\end{equation}
Here, $\mathbf{u}$ is the velocity vector, $p$ is pressure, and $\rho$, $\mu$ are fluid density and dynamic viscosity, respectively.
CFD simulations were performed using ANSYS CFX (ANSYS Inc., USA). The mesh size was refined based on a grid independence study, which demonstrated that pressure drop stabilized with about 100,000 elements. The grid independence study is summarized  in the Supplementary Material.

\subsection{A Semi-Analytical Model for Pancreatic Ductal Pressure Prediction}
While full CFD computations provide the highest accuracy, their computational intensity poses a barrier to routine clinical use \citep{candreva2022current,bluestein2017utilizing}. A simpler method that does not require CFD calculations and can be implemented without the need for simulation infrastructure (CFD software and computing desktop) could accelerate the adoption of this framework in the clinical setting. 
Flow within the PD occurs primarily along its axial direction at low Reynolds numbers and simpler models may be obtained by neglecting the radial and azimuthal flows. Nonetheless, the axial geometric variations—such as strictures or dilations—significantly influence local flow characteristics. To address this, we perform a dimensional reduction from the original three-dimensional Navier–Stokes equations (detailed derivation provided in supplementary material) to derive a quasi-1D formulation explicitly incorporating spatial geometric variations.

Starting from the integral form of the Navier–Stokes equations applied to an axially aligned control volume, we assume axisymmetry and integrate over the duct's cross-sectional area. This yields cross-sectionally averaged mass and momentum equations that explicitly retain geometric parameters, such as cross-sectional area \(A(x,t)\), perimeter \(l_w(x)\), and lateral mass flux \(\dot{m}\) due to physiological secretion. The resulting quasi-1D governing equations are:
\begin{equation}
\begin{cases}
\displaystyle \frac{\partial A}{\partial t} + \frac{\partial Q}{\partial x} = \frac{\dot{m}}{\rho}\,l_w, \\[10pt]
\displaystyle \frac{\partial Q}{\partial t} + \frac{\partial}{\partial x}\left(\langle u^2\rangle A\right)
= -\frac{A}{\rho}\frac{\partial p}{\partial x} + \mu\left\langle\frac{\partial^2 u}{\partial r^2}\right\rangle\frac{A}{\rho},
\end{cases}
\label{eq:quasi1D}
\end{equation}
where $<.>$ denotes radial and azimuthal averaging.  
The first equation describes mass conservation, explicitly incorporating wall secretion effects, while the second one is for the axial momentum balance, including inertial and viscous dissipation terms. To further simplify these equations, we follow established analytical modeling frameworks~\citep{AnalyticalmodelFormaggiabook, AnalyticalmodelFormaggiapaper} :

to introduce the momentum-flux correction factor \(\alpha\) and viscous resistance parameter \(K_R\):
\begin{equation}
\alpha = \frac{\langle u^2\rangle}{\langle u\rangle^2},\,\,\,\,\,\,K_R = \mu\left\langle\frac{\partial^2 u}{\partial r^2}\right\rangle\frac{A}{\rho}.
\end{equation}
For laminar (Poiseuille) flow condition, which is typical in PD, \(K_R = 8\pi\nu = 8\pi\frac{\mu}{\rho}\).
Under steady-state conditions (\(\partial/\partial t=0\)) and considering distributed wall secretion ($\dot{m}$), 
integrating Eq.(\ref{eq:quasi1D}) provides the total flow rate ($Q$) and pressure drop ($\Delta p)$ across the duct:
\begin{equation}
\begin{cases}
\displaystyle \rho Q = \int_x \Delta\dot{M}(x)\,dx, \\[6pt]
\displaystyle \Delta p = -\int_x\left[K_R\frac{\rho Q}{A^2} + \frac{\rho}{A}\frac{d}{dx}\left(\alpha\frac{Q^2}{A}\right)\right]dx,
\end{cases}
\end{equation}
where \(\Delta\dot{M}(x)\) represents local mass inflow per unit duct length. 
To facilitate numerical implementation, the total pressure drop along the pancreatic duct can be succinctly expressed as:
\begin{small}
\begin{equation}
\begin{aligned}
\Delta p = & -K_R\,\dot{m}\int_{x_0}^{x_{\max}}\frac{1}{A^2}\left(\int_{x_0}^{x}dS\right)dx \\
& -\frac{\alpha}{\rho}\,\dot{m}^2\left[
\int_{x_0}^{x_{\max}}\frac{2\,\Delta S(x)}{A^2}\left(\int_{x_0}^{x}dS\right)dx
-\int_{x_0}^{x_{\max}}\frac{\left(\int_{x_0}^{x}dS\right)^2}{A^3}\frac{dA}{dx}\,dx
\right],
\end{aligned}
\label{Dis_Equation}
\end{equation}
\end{small}
where \(\dot{m}\) is the mass flux per unit duct-wall area and acts as the primary independent variable driving the system. The integrals depend solely on the duct geometry, allowing their precomputation for patient-specific cases. 

For computational efficiency and clarity, we define two geometry-dependent integral functions: a nonlinear resistance function \(F_{\mathrm{NL}}\) and a linear resistance function \(F_{\mathrm{L}}\). These integrals explicitly depend on the local cross-sectional area \(A(x)\) and duct surface geometry \(S(x)\), simplifying the pressure drop equation to the following compact form:
\begin{equation}
\Delta p = -\frac{\alpha}{\rho}\,F_{\mathrm{NL}} \left( A(x), S(x)\right) \dot{m}^2 
- K_R\,F_{\mathrm{L}}\left(A(x), S(x) \right) \dot{m}.
\label{eq:dp1D}
\end{equation}
Equivalently, this expression can be further simplified by introducing lumped geometric resistance coefficients \(R\) and \(R_m\):
\begin{equation}
\Delta p = -R\,\dot{m}^2 - R_m\,\dot{m},
\label{eq:RRm}
\end{equation}
where \(R=\alpha F_{\mathrm{NL}}/\rho\) and \(R_m=K_R F_\textrm{L}\) are non-linear and linear resistance, respectively. Note that the resistance coefficients are independent of the mass flow rate. These coefficients can be precomputed from patient-specific geometrical data, enabling rapid and clinically applicable predictions of ductal pressure.

\subsection{Patient-Specific Cases}

In the present study, subsets of patient-specific imaging data collected for the prospective and retrospective cohorts at Johns Hopkins hospital are employed. This study was approved by the Johns Hopkins Medicine Institutional Review Board (IRB No. IRB00427140, approved on August 7, 2024). 

\subsubsection{Prospective Cohort}
For the prospective study, we analyzed 27 patients who had chronic pancreatitis or another pathology and were scheduled to undergo ERCP. On day 1 they underwent MRI MRCP (with or without secretin hormone that stimulates the pancreas juice flow). On day 2 they underwent ERCP with placement of a fractional flow reserve (FFR) pressure sensor guide-wire inside the pancreatic duct and took pressure measurements in the duodenum, papilla, head, body and tail of the pancreas. We then compared the pressure obtained from the MRI MRCP based simulation against the ERCP measured pressures. 

In this prospective cohort of 27 patients, 16 were excluded due to absent, inadequate, or poor-quality MRCP. The remaining 11 underwent successful pancreatic duct reconstruction and CFD-based pressure simulation, as shown in Figure~\ref{fig:group0_pressure}. This subset is referred to as Group 0.

\subsubsection{Retrospective Cohort}
For the retrospective component of the study, adult patients with chronic pancreatitis (CP) who underwent MRI with MRCP at Johns Hopkins Hospital were reviewed. Among this CP cohort, 82 patients had documented clinical pain scores, and 53 of these had high-quality 3D MRCP data suitable for pancreatic duct reconstruction and CFD simulation. This subgroup is referred to as ``Group 1," and was used to examine the relationship between simulated intraductal pressure and patient-reported pain relief score. CFD-simulation results are shown in Figure~\ref{fig:Group1CFDPressure}.

The premise of this analysis is that only PDH-related pain is expected to respond to decompressive interventions such as endoscopic stenting or surgical ductal drainage, while neuropathic pain tends to be refractory. Thus, pain relief after a decompressive procedure should correlate with the magnitude of pressure drop across the length of the pancreatic duct. Thus, a positive correlation between the pain-relief score and the computed pressure drop would provide a validation of the PD simulation pipeline and methodology.
 Correlations between CFD-derived pressure metrics and clinical obtained pain-relief score for this group are shown in Figure~\ref{Painscorecorrelation}. The results of this subgroup analysis are presented in the following sections.

\section{Results}

\subsection{Pressure Distribution in Patient-Specific Pancreatic Ducts}

\begin{figure}[H]
    \centering
    \includegraphics[width=\textwidth]{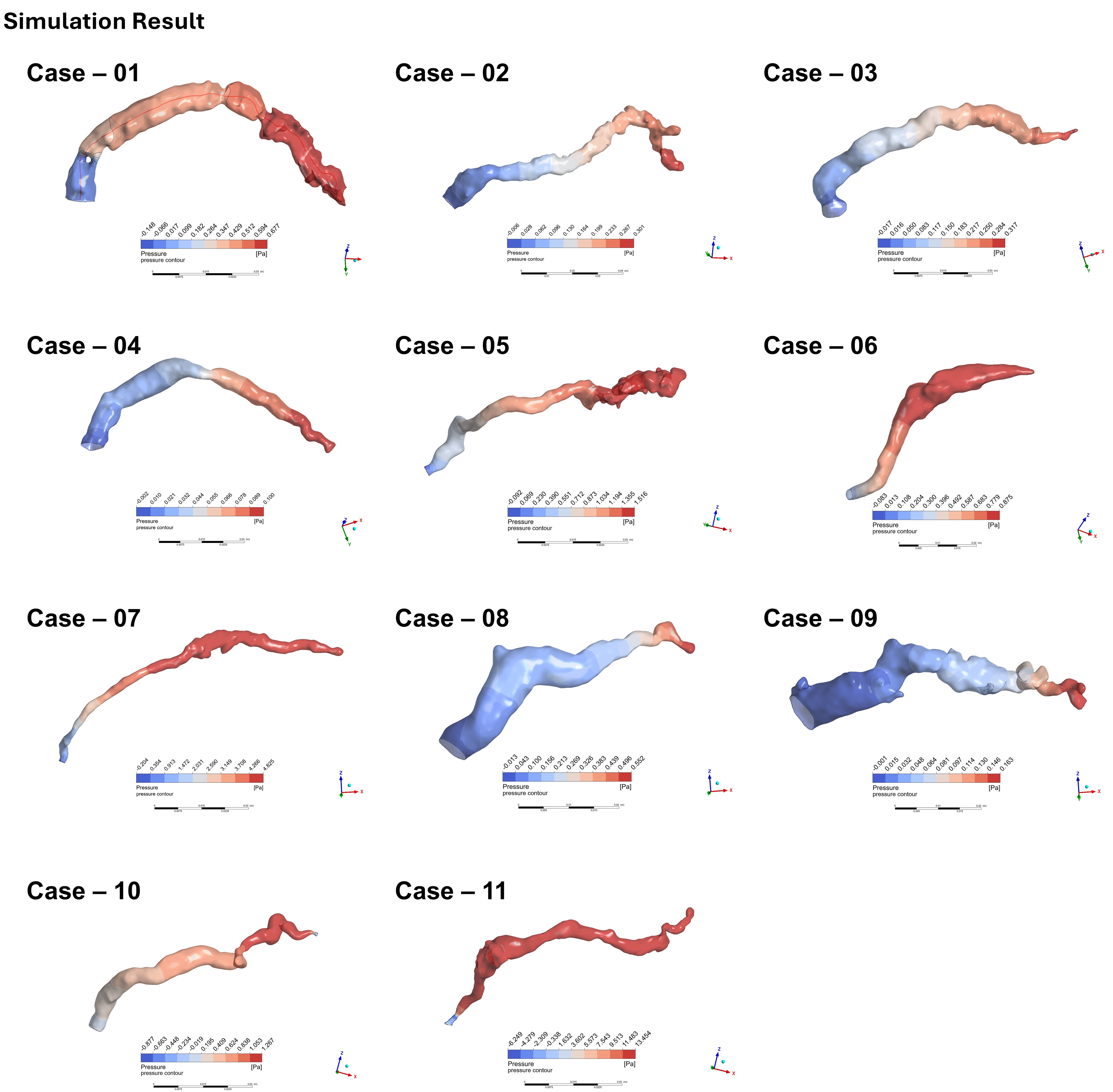
}
    \caption{Pressure distributions in the prospective study cohort (Group 0), CFD-simulated pressure contours: color indicates magnitude.}
    \label{fig:group0_pressure}
\end{figure}

Computational fluid dynamics (CFD) simulations revealed detailed pressure distributions within anatomically realistic pancreatic duct (PD) geometries. These simulations included the prospective study cohort (Group 0) and the retrospective chronic pancreatitis (CP) cohort (Group 1), as illustrated in Figure~\ref{fig:group0_pressure}.
Because the outlet was assigned a Dirichlet boundary condition with zero gauge pressure, the computed pressure gradients primarily reflect cumulative hydraulic resistance and distributed fluid secretion along the duct wall. Consistently, the highest pressures were observed near the pancreatic tail and decreased toward the head (outlet). These trends align with anatomical features such as strictures identified on MRCP, and were further clarified through 3D geometric reconstructions, particularly in regions with subtle or ambiguous narrowing.
Importantly, CFD results enabled precise quantification of the impact of ductal narrowing on local pressure gradients. High-pressure zones consistently co-localized with anatomical strictures, reinforcing their role in contributing to pancreatic ductal hypertension (PDH).

\begin{figure}[H]
    \centering
    \includegraphics[width=\textwidth]{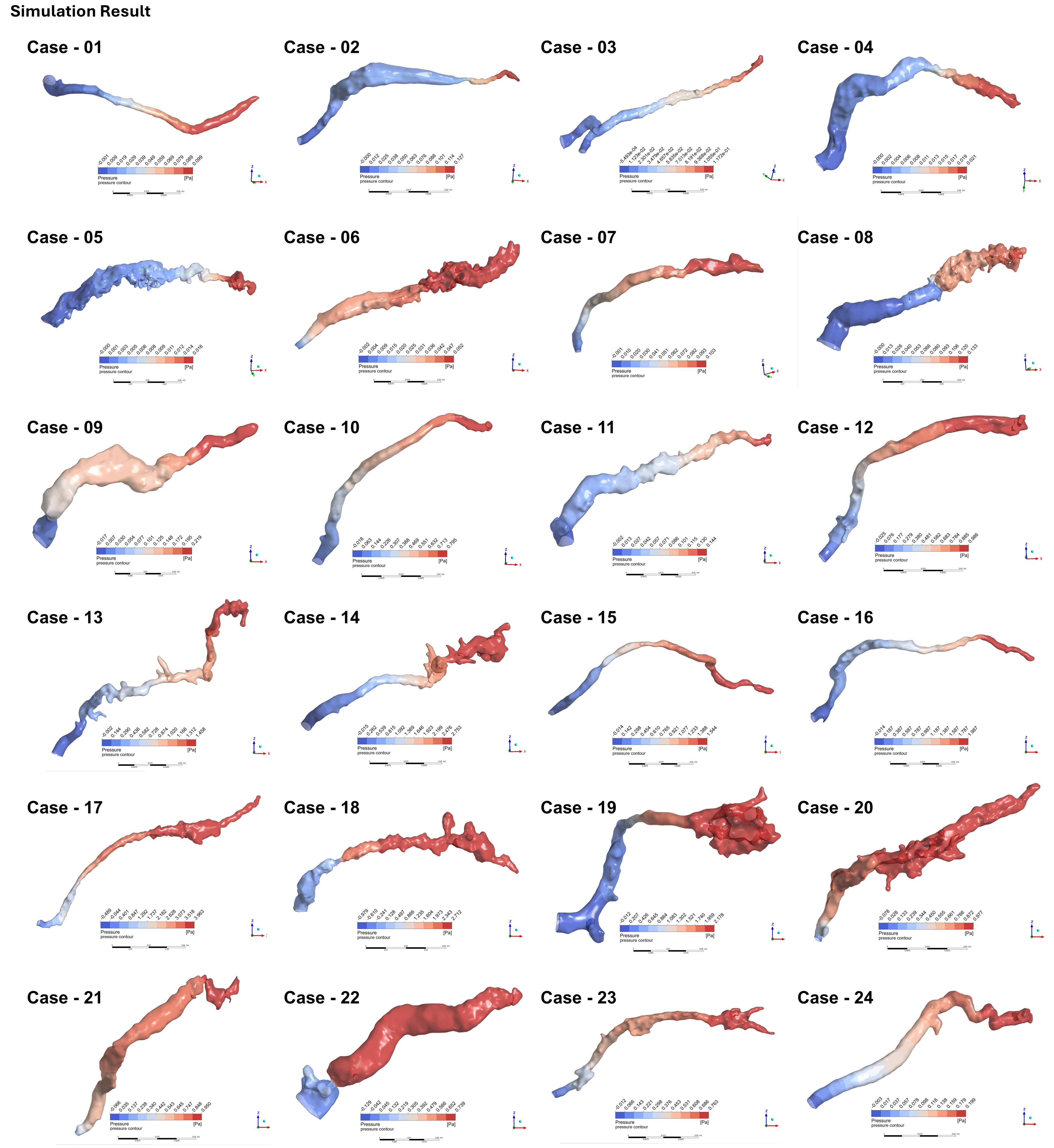}
    \caption{CFD-derived pressure maps for Hopkins CP Group 1, showing localized pressure elevations near anatomical strictures.}
    \label{fig:Group1CFDPressure}
\end{figure}

In contrast, patients in the retrospective CP cohorts (Figures~\ref{fig:Group1CFDPressure}) exhibited more heterogeneous pressure fields, often reflecting multiple and more severe ductal strictures. These regions of anatomical narrowing correlated precisely with elevated pressures in the CFD results, offering direct mechanistic insights into the pathophysiology of PDH.

Inter-patient variability in ductal anatomy resulted in substantial differences in pressure profiles. These findings underscore the importance of personalized CFD modeling in the assessment and management of PDH.

\subsection{Centerline Pressure and Geometric Profiles}

To better understand the mechanistic relationship between ductal morphology and intraductal pressure distribution, we performed a centerline-based analysis of geometric and hemodynamic parameters. This approach enables the localization of strictures and the identification of high-resistance regions based on cross-sectional area and pressure gradients. By translating complex three-dimensional flow patterns into a one-dimensional representation along the ductal centerline, it facilitates clinical interpretation and supports future automation of pressure-risk assessment tools.

Anatomical strictures were algorithmically identified by combining geometric and hemodynamic criteria. Specifically, regions along the centerline exhibiting both cross-sectional area below the 25th percentile and pressure gradient ($\mathrm{d}p/\mathrm{d}x$) steeper than the 10th percentile (more negative) were classified as strictures. These segments were subsequently highlighted in visualizations.
\begin{figure}[H]
    \centering
    \includegraphics[width=\textwidth]{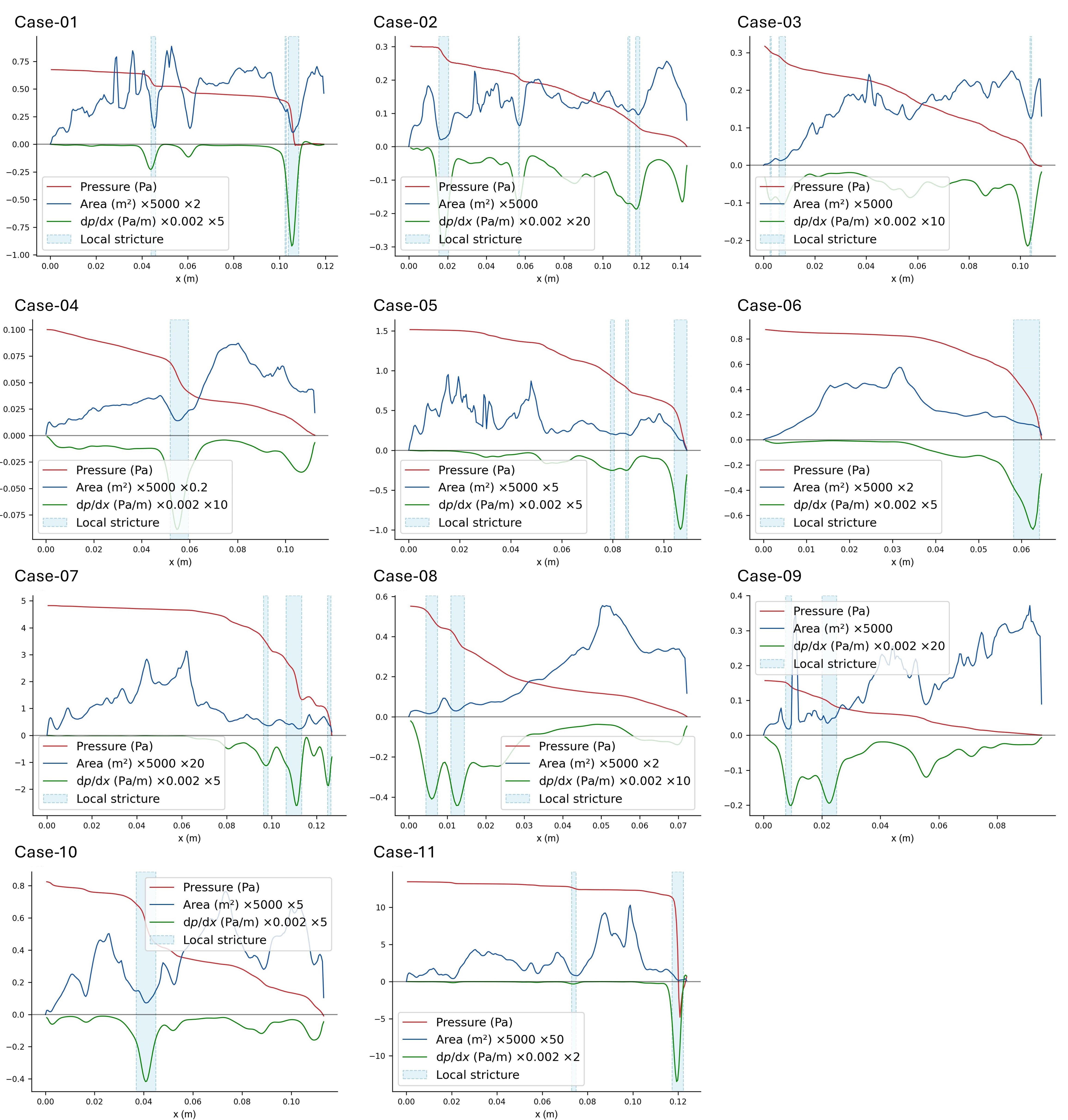}
    \caption{
    Centerline-derived profiles of pressure, cross-sectional area (CSA), and pressure gradient ($\mathrm{d}p/\mathrm{d}x$) in selected Group 0 patients. Pressure peaks consistently align with regions of abrupt CSA reduction, indicating anatomical strictures. Detected stricture regions are highlighted with shaded blue areas.
    }
    \label{fig:pressure_profile_group0}
\end{figure}

As illustrated in Figure~\ref{fig:pressure_profile_group0}, local minima in cross-sectional area (CSA) coincided with sharp pressure drops, demonstrating the functional impact of ductal narrowing on intraductal fluid resistance.

Moreover, regions with larger wall surface areas upstream of strictures produced greater flux under mass source conditions, contributing to additional pressure buildup (Figure~\ref{fig:pressure_profile_group0}, case 27).

\subsection{Validation with ERCP-Pressure Measurement}

Validation against invasive clinical measurements is critical to establishing the credibility of image-based simulations. ERCP remains the only method currently available for direct in vivo measurement of pancreatic ductal pressure, despite its procedural risks. By comparing our MRCP-CFD derived pressure estimates with ERCP-based guidewire measurements, we aim to assess the physiological accuracy of our non-invasive modeling framework and benchmark its diagnostic potential relative to the clinical gold standard.

\begin{figure}[H]
\centering
\includegraphics[width=0.8\textwidth]{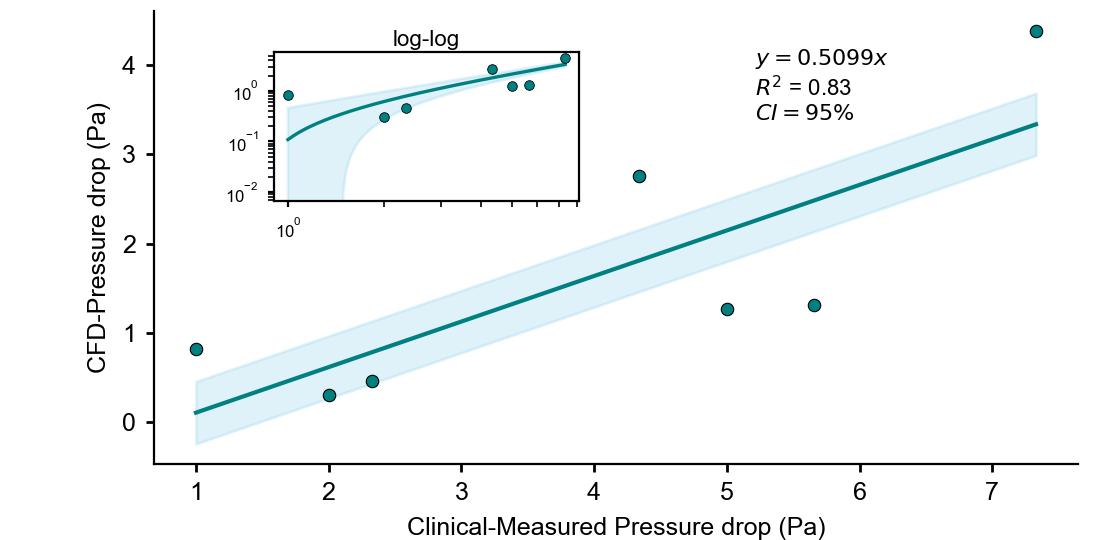}
\caption{Correlation between MRCP-CFD simulated (y-axis) and ERCP-measured (x-axis) pressure drop. All comparison cases are from group 0 patients who underwent secretin-stimulated ERCP. The CFD simulation was performed using a fixed mass inflow rate of \(\dot{m} = 0.1\,\mathrm{g/s}\)}
\label{fig:ercp_correlation}
\end{figure}

All comparison cases were selected from group 0 (prospective cohort) patients who underwent secretin-stimulated ERCP, during which intraductal pressure measurements were recorded. These patients represent the only subset with both secretin administration and clinically measured pressure drops along the pancreatic duct, allowing physiologically relevant comparison with CFD simulations. However, due to the limited availability of such cases, the sample size for this comparison remains small (only 7 cases available to compare). The correlation between the MRCP-CFD derived pressure drop and the ERCP-based measured pressure drop is examined in Fig.\ref{fig:ercp_correlation}. 
Despite of the limited sample size, the positive linear correlation observed supports the accuracy and clinical relevance of our non-invasive simulation approach, though further validation with larger cohorts is warranted.

\subsection{Correlation with Pain Relief}

In chronic pancreatitis, pain may arise from either neuropathic or pressure-driven (PDH) mechanisms. Importantly, only pressure-mediated pain is expected to respond to decompressive interventions. Therefore, the degree of postoperative pain relief can serve as a clinical proxy for the physiological relevance of PDH in a given patient. By examining the relationship between simulated pressure drops and observed pain relief, we aim to evaluate whether our modeling approach can accurately identify patients most likely to benefit from ductal decompression therapy. Pain relief after the therapy was defined as:
\[
\mathbf{Pain}_{\textrm{ref}} = \mathbf{Pain}_{\textrm{before}} - \mathbf{Pain}_{\textrm{after}},
\]
where $\mathbf{Pain}_{\textrm{before}}$ and $\mathbf{Pain}_{\textrm{after}}$ represent clinician-assessed pain scores (scale 1–10) obtained from clinical records before and after decompressive ductal intervention. A higher value of $\mathbf{Pain}_{\textrm{ref}}$ indicates greater symptom improvement and, by extension, a stronger likelihood of pressure-driven ductal hypertension (PDH) being the source of pain.

\begin{figure}[H]
    \centering
    \includegraphics[width=\textwidth]{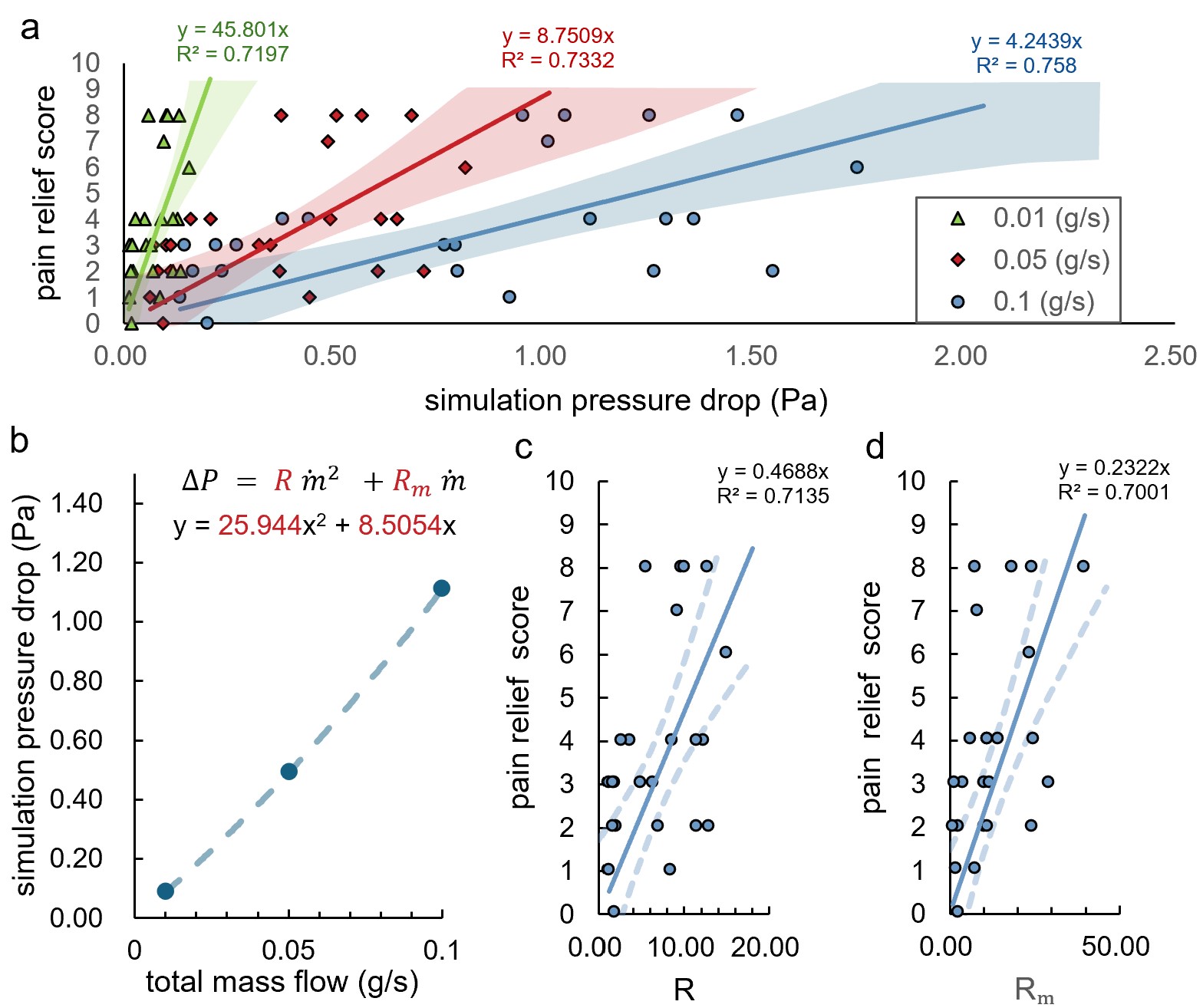}
    \caption{
\textbf{Correlation between simulated pressure drop and clinical pain relief in Group 1 patients.} 
\textbf{a}, Relationship between CFD-simulated pressure drop and clinical pain relief score at three prescribed mass flow rates ($\dot{m} = 0.01$, 0.05, and 0.1~g/s). Linear regression fits are shown with shaded 95\% confidence intervals. 
\textbf{b}, Pressure drop ($\Delta P$) as a function of total mass flow rate, modeled using a quadratic resistance formulation $\Delta P = R \dot{m}^2 + R_m \dot{m}$. 
\textbf{c}, Correlation between pain relief score and fitted linear resistance coefficient $R$. 
\textbf{d}, Correlation between pain relief score and nonlinear resistance coefficient $R_m$. 
Each data point corresponds to an individual subject. $R^2$ values denote the coefficient of determination for each regression.
}
    \label{Painscorecorrelation}
\end{figure}

This study is performed with the group 1 (retrospective cohort) data. The CFD simulations are performed with multiple pancreatic juice mass flow rates ($\dot{m} = 0.01$, 0.05, and 0.1~g/s) to cover wide range of pathological conditions presented in the retrospective cohort. 
The correlations between the simulated pressure drop and the pain relief score are examined in Fig. \ref{Painscorecorrelation}.
As one can see in the figure, the simulated pressure drop was positively correlated with pain relief score across all tested mass flow rates, with coefficients of determination ($R^2$) ranging from 0.72 to 0.76. This supports the hypothesis that higher intraductal pressure gradients are associated with more clinically meaningful symptom relief following decompression therapy.

Further regression analysis using a quadratic resistance model, Eq.(\ref{eq:RRm}), also performed.
The mass flow rate independent linear ($R_m$) and non-linear ($R$) ductal resistances are evaluated by fitting the simulation data to the quadratic resistance model as shown in Fig.\ref{Painscorecorrelation}b for an example.
The correlations between the resistance coefficients and the pain relief score are then examined in Fig.\ref{Painscorecorrelation}c\& d.
The results demonstrated strong agreement between model-predicted resistance coefficients and pain relief outcomes, reinforcing the utility of computational pressure estimates as a potential biomarker.

\subsection{Validation of the 1D Analytical Pressure Drop Model}

\begin{figure}[H]
    \centering
    \includegraphics[width=0.9\textwidth]{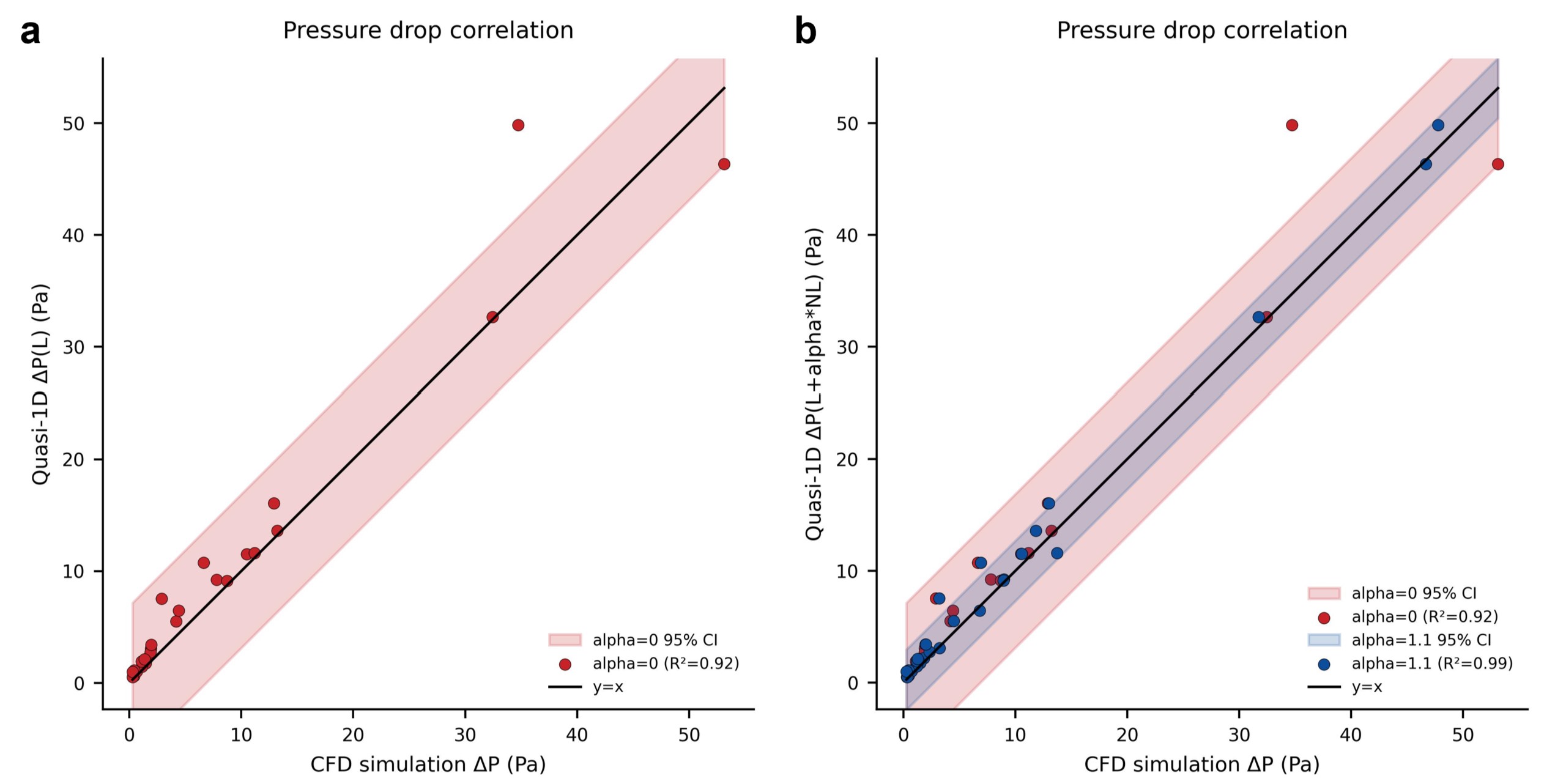
}
    \caption{
    Comparison between CFD-derived and quasi-1D analytical pressure drops among group 0 and group 1 patients.
    (a) Model using only the linear resistance component (\(\alpha = 0\)) shows good correlation with CFD results (\(R^2 = 0.92\)), with deviations appearing at higher pressure gradients.
    (b) Inclusion of a nonlinear inertial loss term (\(\alpha = 1.1\)) substantially improves the agreement (\(R^2 = 0.99\)), particularly in regimes of elevated flow resistance. Shaded regions denote the 95\% confidence intervals. The black diagonal line represents perfect agreement (\(y = x\)).}
    \label{fig:sim_vs_analytic}
\end{figure}

To assess the accuracy of the proposed quasi-one-dimensional (quasi-1D) analytical model, we compared its pressure drop predictions with those obtained from full computational fluid dynamics (CFD) simulations across a range of flow conditions.
For all the cases, the pressure drop across the PD is calculated by using the proposed quasi-1D model, Eq. (\ref{eq:dp1D}) with and without the non-linear inertial loss term ($\alpha$). 
The correlations between the pressure drop obtained by full CFD and quasi-1D model are examined in Fig.\ref{fig:sim_vs_analytic}.

As shown in Figure~\ref{fig:sim_vs_analytic}a, even when only the linear pressure loss term is considered, the quasi-1D model achieves a high degree of accuracy, with a coefficient of determination of \(R^2 = 0.92\). This highlights the model's capability to capture dominant pressure loss behavior under moderate flow conditions.
Incorporating a nonlinear correction factor (\(\alpha = 1.1\), based on pervious research result (\cite{AnalyticalmodelFormaggiapaper}, \cite{smith2002anatomically}) further improves predictive accuracy, particularly in high-flow regimes where inertial losses become significant (Figure~\ref{fig:sim_vs_analytic}b). The nonlinear model yields an excellent agreement with CFD results across the full pressure drop range (\(R^2 = 0.99\)).
These findings validate the quasi-1D analytical model as a computationally efficient and reliable alternative to full CFD simulations. Its balance of speed and accuracy renders it a strong candidate for clinical applications requiring rapid yet robust hemodynamic estimation, such as intraoperative planning or device evaluation.

\section{Discussion}


This study provides the initial quantitative validation of MRCP-CFD against ERCP pressure measurements and demonstrates its clinical relevance through correlation with pain relief following ductal decompression, demonstrates that MRCP-based CFD provides a non-invasive and robust alternative to ERCP-guided intraductal manometry, avoiding risks such as pancreatitis, ductal injury, and catheter-related limitations in complex ductal anatomies reported in previous research (\cite{chandrasekhara2017adverse}).

MRCP yields high-resolution, contrast-free ductal images which, when coupled with CFD, enable generation of continuous intraductal pressure maps, thereby offering functional assessment beyond conventional structural imaging. While conventional MRCP permits morphological assessment of strictures, its diagnostic utility remains limited by its qualitative nature (\cite{goldfinger2020QuantMRCP}). 

The geometric severity of a ductal stricture can significantly influence the internal flow and associated pressure drop along the duct, as shown in early fluid dynamics studies (\cite{mates1978fluid}, \cite{ComputationalModelingofPancreaticDuctFluidFlow_previous}
). This relationship is inherently nonlinear, as demonstrated in recent analyses of bile duct flow (\cite{peng2024analysis}), indicating that even minor additional narrowing can lead to a disproportionately large increase in pressure gradient. This underscores the importance of precise, quantitative evaluation, as the overall outcome is highly sensitive to the accurate characterization of the affected region. Unlike prior pancreatic duct studies that primarily relied on morphological features or surrogate imaging markers (\cite{matos1997pancreatic}, \cite{chen2020prediction}, \cite{takayanagi2023clinical}), our approach integrates anatomy and hemodynamics to provide a direct functional assessment of ductal hypertension.

More importantly, according to experimental studies using 3D-printed ducts (\cite{modi2018optimal}), When there are intertwined lesions (two or more stenoses in the same duct), the results will be interrelated, often underestimating the true severity of a single stenosis. 

MRCP-CFD procedure advances beyond static interpretation by coupling image-derived anatomy with flow simulations, thereby enabling dynamic evaluation of PDH. Using centerline-based cross-sectional profiling and pressure mapping, we identified focal strictures with biomechanical significance—i.e., those responsible for steep pressure gradients and elevated upstream resistance. This dual structural-functional assessment offers a more nuanced framework for interventional planning, while also aligns with the trend toward quantitative MRCP, which enables more precise analyses through advanced visualization techniques (\cite{goldfinger2020QuantMRCP}).

We validated our computational estimates against ERCP-derived intraductal pressure measurements, demonstrating moderate-to-strong agreement across patients. Importantly, a consistent underestimation (~50\%) of absolute pressure was observed. This systematic bias likely reflects three compounding factors: (1) assumed pancreatic secretion rates that may not reflect in vivo variability; (2) modeling simplifications such as rigid wall boundaries and laminar, steady-state flow; and (3) MRCP resolution constraints that prevent accurate capture of micro-strictures or duct wall compliance, in line with previous studies reporting the existence of systematic underestimation in pressure-drop prediction (\cite{nolte2021validation}, \cite{falahatpisheh2016simplified}, \cite{qiu2025numerical}). Despite this systematic underestimation, the relative ranking of patients and the correlation with clinical outcomes remain robust, suggesting that MRCP-CFD can still serve as a valuable stratification tool even without absolute pressure calibration

In our model, the flow contribution from small branches was represented by mass sources imposed at the boundary, as these fine branches—which generate inflow into the pancreatic duct—are difficult to capture using MRCP. Previous studies have retained small branches explicitly and performed simulations with full PD models. Similar strategies have been applied in cerebrospinal fluid research (\cite{vandenbulcke2022computational}) and in biliary fluid dynamics (\cite{meyer2017predictive}). Our approach, as a pragmatic substitute for highly detailed branch modeling, allowed us to reconstruct near-physiological ductal flow boundary conditions.

Our model assumes incompressible, Newtonian fluid behavior within rigid, non‑deformable ducts under steady‑state conditions—which can be considered physiologically reasonable for mild to moderate chronic pancreatitis, where ductal fibrosis is limited and pseudocysts are absent. Previous histopathological and imaging studies confirm that chronic pancreatitis is marked by increased periductal connective tissue and alterations in duct morphology—such as strictures, irregular dilation, and pseudocyst formation—that may significantly alter duct compliance and flow dynamics in advanced disease (\cite{yamashita2022imaging}) Moreover, under conditions of intermittent secretion or duodenal reflux—particularly when protective mechanisms are compromised—unsteady or reversed flow may prevail, as suggested by fluid mechanics modeling of pancreatic juice reflux in congenital anomalies and ductal pressure disruptions (\cite{tajikawa2023investigation}).

Similar simplifications were also adopted in earlier COMSOL-based pancreatic duct CFD studies (\cite{ComputationalModelingofPancreaticDuctFluidFlow_previous}), which used parameterized ductal geometries to evaluate pressure–flow relationships under steady laminar conditions. While such models demonstrated the feasibility of pancreatic duct CFD, they did not incorporate patient-specific imaging data or transient flow effects, thereby limiting their direct clinical applicability. Building on and improving these methodologies, our patient-specific modeling offers a clinically translatable advance by directly linking imaging-derived ductal morphology with functional pressure estimates.

Future work should integrate wall elasticity, non-Newtonian rheology, and transient flow dynamics. 
Vascular biomechanics offers practical strategies via fluid–structure interaction (FSI) to capture deformability and pulsatility; for example, \cite{qiao2019numerical} coupled non-Newtonian FSI with Windkessel outlets, and \cite{kanyanta2009validation} validated a two-way FSI model for compliant conduits. 
Comparative studies further show that elastic-wall, pulsatile models better reproduce realistic hemodynamics than rigid-wall counterparts \cite{athani2021two}.

Beyond methodological considerations, our results carry important clinical implications. In particular, the integration of automated MRCP segmentation and rapid quasi-1D modeling could enable near real-time pressure estimation at the point of care. The observed correlation between simulated pressure drops and symptomatic improvement supports the notion that pancreatic ductal hypertension (PDH) is a distinct pain mechanism in chronic pancreatitis and may be used to stratify patients for endoscopic or surgical decompression. By quantifying duct-specific resistance and identifying focal regions of maximal pressure gradient, MRCP-CFD offers a functional complement to morphological imaging that can guide targeted interventions. At the same time, the modest sample size, dependence on MRCP image quality, and simplifying assumptions such as rigid walls and steady flow highlight the need for further validation in larger, prospective cohorts. Future developments—such as incorporating fluid-structure interaction, non-Newtonian rheology, and automated image-based modeling—may enable real-time clinical integration of this approach. Taken together, these findings establish MRCP-CFD as a promising step toward non-invasive, functionally informed diagnosis of pancreatic ductal hypertension in chronic pancreatitis

\section*{Author Contributions}
\textbf{Haobo Zhao:} Methodology,  Validation, Formal Analysis, Investigation, Data Curation, Writing – Original Draft, Visualization
\textbf{Jung-Hee Seo:} Conceptualization, Methodology, Software, Supervision, Project Administration, Writing – Original Draft \& Review \& Editing
\textbf{Venkata Akshintala:} Conceptualization, Supervision, Project Administration, Resources, Writing – Review \& Editing
\textbf{Surya Evani:} Investigation, Resources
\textbf{Rajat Mittal:} Conceptualization, Supervision, Project Administration, Writing – Review \& Editing, Funding Acquisition

\section*{Supplementary Information}
Supplementary material is available online, including:
\begin{itemize}
  \item Grid independence study (Section S1)
  \item Derivation of quasi-one-dimensional pressure model (Section S2)
  \item Additional validation plots comparing CFD and analytical models (Section S3)
\end{itemize}

\section*{Funding}
This work was supported by Johns Hopkins University School of Medicine, Division of Gastroenterology, Pilot Grant, National Pancreas Foundation Grant.

\section*{Conflicts of Interest}
The authors declare that they have no conflict of interest.

\section*{Ethics Approval and Consent to Participate}
The study was approved by the Johns Hopkins Medicine Institutional Review Board (IRB No. IRB00427140, approved on August 7, 2024). All participants provided informed consent for data collection and study participation. 
Clinical trial number: not applicable.

\section*{Consent for Publication}
All authors have read and approved the final version of the manuscript and consent to its publication.

\section*{Data Availability}
The datasets generated and/or analyzed during the current study are not publicly available due to patient privacy restrictions, but are available from the corresponding author on reasonable request.

\section*{Code Availability}
The custom computational code used for the quasi-one-dimensional model and post-processing is available from the corresponding author upon reasonable request.


\bibliography{sn-bibliography}

\begin{thebibliography}{}
\renewcommand{\doi}[1]{\url{https://doi.org/#1}}
\bibcommenthead

\bibitem [\protect \citeauthoryear {%
Athani%
\ \protect \BOthers {.}}{%
Athani%
\ \protect \BOthers {.}}{%
{\protect \APACyear {2021}}%
}]{%
athani2021two}
\APACinsertmetastar {%
athani2021two}%
\begin{APACrefauthors}%
Athani, A.%
, Ghazali, N.N.N.%
, Badruddin, I.A.%
, Usmani, A.Y.%
, Kamangar, S.%
, Anqi, A.E.%
\BCBL {} Ahammad, N.A.%
\end{APACrefauthors}%
\unskip\
\newblock
\APACrefYearMonthDay{2021}{}{}.
\newblock
{\BBOQ}\APACrefatitle {Two-phase non-Newtonian pulsatile blood flow simulations
  in a rigid and flexible patient-specific left coronary artery (LCA)
  exhibiting multi-stenosis} {Two-phase non-newtonian pulsatile blood flow
  simulations in a rigid and flexible patient-specific left coronary artery
  (lca) exhibiting multi-stenosis}.{\BBCQ}
\newblock
\APACjournalVolNumPages{Applied Sciences}{11}{23}{11361,}
\newblock

\newblock

\PrintBackRefs{\CurrentBib}

\bibitem [\protect \citeauthoryear {%
Atzeni%
\ \protect \BOthers {.}}{%
Atzeni%
\ \protect \BOthers {.}}{%
{\protect \APACyear {2021}}%
}]{%
atzeni2021computational}
\APACinsertmetastar {%
atzeni2021computational}%
\begin{APACrefauthors}%
Atzeni, C.%
, Lesma, G.%
, Dubini, G.%
, Masi, M.%
, Rossi, F.%
\BCBL {} Bianchi, E.%
\end{APACrefauthors}%
\unskip\
\newblock
\APACrefYearMonthDay{2021}{}{}.
\newblock
{\BBOQ}\APACrefatitle {Computational fluid dynamic models as tools to predict
  aerosol distribution in tracheobronchial airways} {Computational fluid
  dynamic models as tools to predict aerosol distribution in tracheobronchial
  airways}.{\BBCQ}
\newblock
\APACjournalVolNumPages{Scientific Reports}{11}{1}{1109,}
\newblock

\newblock

\PrintBackRefs{\CurrentBib}

\bibitem [\protect \citeauthoryear {%
Bali%
\ \protect \BOthers {.}}{%
Bali%
\ \protect \BOthers {.}}{%
{\protect \APACyear {2005}}%
}]{%
bali2005quantification}
\APACinsertmetastar {%
bali2005quantification}%
\begin{APACrefauthors}%
Bali, M.A.%
, Sztantics, A.%
, Metens, T.%
, Arvanitakis, M.%
, Delhaye, M.%
, Devi{\`e}re, J.%
\BCBL {} Matos, C.%
\end{APACrefauthors}%
\unskip\
\newblock
\APACrefYearMonthDay{2005}{}{}.
\newblock
{\BBOQ}\APACrefatitle {Quantification of pancreatic exocrine function with
  secretin-enhanced magnetic resonance cholangiopancreatography: normal values
  and short-term effects of pancreatic duct drainage procedures in chronic
  pancreatitis. Initial results} {Quantification of pancreatic exocrine
  function with secretin-enhanced magnetic resonance cholangiopancreatography:
  normal values and short-term effects of pancreatic duct drainage procedures
  in chronic pancreatitis. initial results}.{\BBCQ}
\newblock
\APACjournalVolNumPages{European radiology}{15}{}{2110--2121,}
\newblock

\newblock

\PrintBackRefs{\CurrentBib}

\bibitem [\protect \citeauthoryear {%
Beyer%
, Habtezion%
, Werner%
, Lerch%
\BCBL {}\ \BBA {} Mayerle%
}{%
Beyer%
\ \protect \BOthers {.}}{%
{\protect \APACyear {2020}}%
}]{%
beyer2020chronic}
\APACinsertmetastar {%
beyer2020chronic}%
\begin{APACrefauthors}%
Beyer, G.%
, Habtezion, A.%
, Werner, J.%
, Lerch, M.M.%
\BCBL {} Mayerle, J.%
\end{APACrefauthors}%
\unskip\
\newblock
\APACrefYearMonthDay{2020}{}{}.
\newblock
{\BBOQ}\APACrefatitle {Chronic pancreatitis} {Chronic pancreatitis}.{\BBCQ}
\newblock
\APACjournalVolNumPages{The Lancet}{396}{10249}{499--512,}
\newblock

\newblock

\PrintBackRefs{\CurrentBib}

\bibitem [\protect \citeauthoryear {%
Bluestein%
}{%
Bluestein%
}{%
{\protect \APACyear {2017}}%
}]{%
bluestein2017utilizing}
\APACinsertmetastar {%
bluestein2017utilizing}%
\begin{APACrefauthors}%
Bluestein, D.%
\end{APACrefauthors}%
\unskip\
\newblock
\APACrefYearMonthDay{2017}{}{}.
\newblock
{\BBOQ}\APACrefatitle {Utilizing Computational Fluid Dynamics in Cardiovascular
  Engineering and Medicine-What You Need to Know. Its Translation to the
  Clinic/Bedside} {Utilizing computational fluid dynamics in cardiovascular
  engineering and medicine-what you need to know. its translation to the
  clinic/bedside}.{\BBCQ}
\newblock
\APACjournalVolNumPages{Artificial organs}{41}{2}{117--121,}
\newblock
\begin{APACrefDOI} \doi{10.1111/aor.12914} \end{APACrefDOI}
\newblock

\newblock

\PrintBackRefs{\CurrentBib}

\bibitem [\protect \citeauthoryear {%
Candreva%
\ \protect \BOthers {.}}{%
Candreva%
\ \protect \BOthers {.}}{%
{\protect \APACyear {2022}}%
}]{%
candreva2022current}
\APACinsertmetastar {%
candreva2022current}%
\begin{APACrefauthors}%
Candreva, A.%
, De~Nisco, G.%
, Rizzini, M.L.%
, D’Ascenzo, F.%
, De~Ferrari, G.M.%
, Gallo, D.%
\BDBL {}Chiastra, C.%
\end{APACrefauthors}%
\unskip\
\newblock
\APACrefYearMonthDay{2022}{}{}.
\newblock
{\BBOQ}\APACrefatitle {Current and future applications of computational fluid
  dynamics in coronary artery disease} {Current and future applications of
  computational fluid dynamics in coronary artery disease}.{\BBCQ}
\newblock
\APACjournalVolNumPages{Reviews in Cardiovascular Medicine}{23}{11}{377,}
\newblock

\newblock

\PrintBackRefs{\CurrentBib}

\bibitem [\protect \citeauthoryear {%
Ceyhan%
\ \protect \BOthers {.}}{%
Ceyhan%
\ \protect \BOthers {.}}{%
{\protect \APACyear {2009}}%
}]{%
ceyhan2009pancreatic}
\APACinsertmetastar {%
ceyhan2009pancreatic}%
\begin{APACrefauthors}%
Ceyhan, G.O.%
, Bergmann, F.%
, Kadihasanoglu, M.%
, Altintas, B.%
, Demir, I.E.%
, Hinz, U.%
\BDBL {}others%
\end{APACrefauthors}%
\unskip\
\newblock
\APACrefYearMonthDay{2009}{}{}.
\newblock
{\BBOQ}\APACrefatitle {Pancreatic neuropathy and neuropathic pain—a
  comprehensive pathomorphological study of 546 cases} {Pancreatic neuropathy
  and neuropathic pain—a comprehensive pathomorphological study of 546
  cases}.{\BBCQ}
\newblock
\APACjournalVolNumPages{Gastroenterology}{136}{1}{177--186,}
\newblock

\newblock

\PrintBackRefs{\CurrentBib}

\bibitem [\protect \citeauthoryear {%
Chandrasekhara%
\ \protect \BOthers {.}}{%
Chandrasekhara%
\ \protect \BOthers {.}}{%
{\protect \APACyear {2017}}%
}]{%
chandrasekhara2017adverse}
\APACinsertmetastar {%
chandrasekhara2017adverse}%
\begin{APACrefauthors}%
Chandrasekhara, V.%
, Khashab, M.A.%
, Muthusamy, V.R.%
, Acosta, R.D.%
, Agrawal, D.%
, Bruining, D.H.%
\BDBL {}others%
\end{APACrefauthors}%
\unskip\
\newblock
\APACrefYearMonthDay{2017}{}{}.
\newblock
{\BBOQ}\APACrefatitle {Adverse events associated with ERCP} {Adverse events
  associated with ercp}.{\BBCQ}
\newblock
\APACjournalVolNumPages{Gastrointestinal endoscopy}{85}{1}{32--47,}
\newblock

\newblock

\PrintBackRefs{\CurrentBib}

\bibitem [\protect \citeauthoryear {%
Chen%
\ \protect \BOthers {.}}{%
Chen%
\ \protect \BOthers {.}}{%
{\protect \APACyear {2020}}%
}]{%
chen2020prediction}
\APACinsertmetastar {%
chen2020prediction}%
\begin{APACrefauthors}%
Chen, W.%
, Butler, R.K.%
, Zhou, Y.%
, Parker, R.A.%
, Jeon, C.Y.%
\BCBL {} Wu, B.U.%
\end{APACrefauthors}%
\unskip\
\newblock
\APACrefYearMonthDay{2020}{}{}.
\newblock
{\BBOQ}\APACrefatitle {Prediction of pancreatic cancer based on imaging
  features in patients with duct abnormalities} {Prediction of pancreatic
  cancer based on imaging features in patients with duct abnormalities}.{\BBCQ}
\newblock
\APACjournalVolNumPages{Pancreas}{49}{3}{413--419,}
\newblock

\newblock

\PrintBackRefs{\CurrentBib}

\bibitem [\protect \citeauthoryear {%
DuVal~Jr%
}{%
DuVal~Jr%
}{%
{\protect \APACyear {1958}}%
}]{%
duval1958effect}
\APACinsertmetastar {%
duval1958effect}%
\begin{APACrefauthors}%
DuVal~Jr, M.%
\end{APACrefauthors}%
\unskip\
\newblock
\APACrefYearMonthDay{1958}{}{}.
\newblock
{\BBOQ}\APACrefatitle {The effect of chronic pancreatitis on pressure tolerance
  in the human pancreatic duct.} {The effect of chronic pancreatitis on
  pressure tolerance in the human pancreatic duct.}{\BBCQ}
\newblock
\APACjournalVolNumPages{Surgery}{43}{5}{798--801,}
\newblock

\newblock

\PrintBackRefs{\CurrentBib}

\bibitem [\protect \citeauthoryear {%
Falahatpisheh%
\ \protect \BOthers {.}}{%
Falahatpisheh%
\ \protect \BOthers {.}}{%
{\protect \APACyear {2016}}%
}]{%
falahatpisheh2016simplified}
\APACinsertmetastar {%
falahatpisheh2016simplified}%
\begin{APACrefauthors}%
Falahatpisheh, A.%
, Rickers, C.%
, Gabbert, D.%
, Heng, E.L.%
, Stalder, A.%
, Kramer, H\BHBI H.%
\BDBL {}Kheradvar, A.%
\end{APACrefauthors}%
\unskip\
\newblock
\APACrefYearMonthDay{2016}{}{}.
\newblock
{\BBOQ}\APACrefatitle {Simplified Bernoulli's method significantly
  underestimates pulmonary transvalvular pressure drop} {Simplified bernoulli's
  method significantly underestimates pulmonary transvalvular pressure
  drop}.{\BBCQ}
\newblock
\APACjournalVolNumPages{Journal of Magnetic Resonance
  Imaging}{43}{6}{1313--1319,}
\newblock

\newblock

\PrintBackRefs{\CurrentBib}

\bibitem [\protect \citeauthoryear {%
Formaggia%
, Lamponi%
\BCBL {}\ \BBA {} Quarteroni%
}{%
Formaggia%
\ \protect \BOthers {.}}{%
{\protect \APACyear {2003}}%
}]{%
AnalyticalmodelFormaggiapaper}
\APACinsertmetastar {%
AnalyticalmodelFormaggiapaper}%
\begin{APACrefauthors}%
Formaggia, L.%
, Lamponi, D.%
\BCBL {} Quarteroni, A.%
\end{APACrefauthors}%
\unskip\
\newblock
\APACrefYearMonthDay{2003}{}{}.
\newblock
{\BBOQ}\APACrefatitle {One-dimensional models for blood flow in arteries}
  {One-dimensional models for blood flow in arteries}.{\BBCQ}
\newblock
\APACjournalVolNumPages{Journal of engineering mathematics}{47}{}{251--276,}
\newblock

\newblock

\PrintBackRefs{\CurrentBib}

\bibitem [\protect \citeauthoryear {%
Formaggia%
, Quarteroni%
\BCBL {}\ \BBA {} Veneziani%
}{%
Formaggia%
\ \protect \BOthers {.}}{%
{\protect \APACyear {2010}}%
}]{%
AnalyticalmodelFormaggiabook}
\APACinsertmetastar {%
AnalyticalmodelFormaggiabook}%
\begin{APACrefauthors}%
Formaggia, L.%
, Quarteroni, A.%
\BCBL {} Veneziani, A.%
\end{APACrefauthors}%
\unskip\
\newblock
\APACrefYear{2010}.
\newblock
\APACrefbtitle {Cardiovascular Mathematics: Modeling and simulation of the
  circulatory system} {Cardiovascular mathematics: Modeling and simulation of
  the circulatory system}\ (\BVOL~1).
\newblock
\APACaddressPublisher{}{Springer Science \& Business Media}.
\PrintBackRefs{\CurrentBib}

\bibitem [\protect \citeauthoryear {%
Goldfinger%
\ \protect \BOthers {.}}{%
Goldfinger%
\ \protect \BOthers {.}}{%
{\protect \APACyear {2020}}%
}]{%
goldfinger2020QuantMRCP}
\APACinsertmetastar {%
goldfinger2020QuantMRCP}%
\begin{APACrefauthors}%
Goldfinger, M.H.%
, Ridgway, G.R.%
, Ferreira, C.%
, Langford, C.R.%
, Cheng, L.%
, Kazimianec, A.%
\BDBL {}others%
\end{APACrefauthors}%
\unskip\
\newblock
\APACrefYearMonthDay{2020}{}{}.
\newblock
{\BBOQ}\APACrefatitle {Quantitative MRCP imaging: accuracy, repeatability,
  reproducibility, and cohort-derived normative ranges} {Quantitative mrcp
  imaging: accuracy, repeatability, reproducibility, and cohort-derived
  normative ranges}.{\BBCQ}
\newblock
\APACjournalVolNumPages{Journal of Magnetic Resonance
  Imaging}{52}{3}{807--820,}
\newblock

\newblock

\PrintBackRefs{\CurrentBib}

\bibitem [\protect \citeauthoryear {%
Kanyanta%
, Ivankovic%
\BCBL {}\ \BBA {} Karac%
}{%
Kanyanta%
\ \protect \BOthers {.}}{%
{\protect \APACyear {2009}}%
}]{%
kanyanta2009validation}
\APACinsertmetastar {%
kanyanta2009validation}%
\begin{APACrefauthors}%
Kanyanta, V.%
, Ivankovic, A.%
\BCBL {} Karac, A.%
\end{APACrefauthors}%
\unskip\
\newblock
\APACrefYearMonthDay{2009}{}{}.
\newblock
{\BBOQ}\APACrefatitle {Validation of a fluid--structure interaction numerical
  model for predicting flow transients in arteries} {Validation of a
  fluid--structure interaction numerical model for predicting flow transients
  in arteries}.{\BBCQ}
\newblock
\APACjournalVolNumPages{Journal of Biomechanics}{42}{11}{1705--1712,}
\newblock

\newblock

\PrintBackRefs{\CurrentBib}

\bibitem [\protect \citeauthoryear {%
Kleeff%
\ \protect \BOthers {.}}{%
Kleeff%
\ \protect \BOthers {.}}{%
{\protect \APACyear {2017}}%
}]{%
kleeff2017chronic}
\APACinsertmetastar {%
kleeff2017chronic}%
\begin{APACrefauthors}%
Kleeff, J.%
, Whitcomb, D.C.%
, Shimosegawa, T.%
, Esposito, I.%
, Lerch, M.M.%
, Gress, T.%
\BDBL {}others%
\end{APACrefauthors}%
\unskip\
\newblock
\APACrefYearMonthDay{2017}{}{}.
\newblock
{\BBOQ}\APACrefatitle {Chronic pancreatitis} {Chronic pancreatitis}.{\BBCQ}
\newblock
\APACjournalVolNumPages{Nature reviews Disease primers}{3}{1}{1--18,}
\newblock

\newblock

\PrintBackRefs{\CurrentBib}

\bibitem [\protect \citeauthoryear {%
Lechelek%
, Horna%
, Zrour%
, Naudin%
\BCBL {}\ \BBA {} Guillevin%
}{%
Lechelek%
\ \protect \BOthers {.}}{%
{\protect \APACyear {2022}}%
}]{%
lechelek2022hybrid}
\APACinsertmetastar {%
lechelek2022hybrid}%
\begin{APACrefauthors}%
Lechelek, L.%
, Horna, S.%
, Zrour, R.%
, Naudin, M.%
\BCBL {} Guillevin, C.%
\end{APACrefauthors}%
\unskip\
\newblock
\APACrefYearMonthDay{2022}{}{}.
\newblock
{\BBOQ}\APACrefatitle {A hybrid method for 3d reconstruction of mr images} {A
  hybrid method for 3d reconstruction of mr images}.{\BBCQ}
\newblock
\APACjournalVolNumPages{Journal of Imaging}{8}{4}{103,}
\newblock

\newblock

\PrintBackRefs{\CurrentBib}

\bibitem [\protect \citeauthoryear {%
Li%
\ \protect \BOthers {.}}{%
Li%
\ \protect \BOthers {.}}{%
{\protect \APACyear {2024}}%
}]{%
li2024impact}
\APACinsertmetastar {%
li2024impact}%
\begin{APACrefauthors}%
Li, X.%
, Ni, X.%
, Sun, W.%
, Liu, J.%
, Shang, Y.%
, Liu, H.%
\BCBL {} Tu, J.%
\end{APACrefauthors}%
\unskip\
\newblock
\APACrefYearMonthDay{2024}{}{}.
\newblock
{\BBOQ}\APACrefatitle {The impact of choledochal cysts on bile fluid dynamics:
  A perspective using computational fluid dynamics and surface mapping
  technique} {The impact of choledochal cysts on bile fluid dynamics: A
  perspective using computational fluid dynamics and surface mapping
  technique}.{\BBCQ}
\newblock
\APACjournalVolNumPages{Physics of Fluids}{36}{6}{,}
\newblock

\newblock

\PrintBackRefs{\CurrentBib}

\bibitem [\protect \citeauthoryear {%
Madsen%
\ \BBA {} Winkler%
}{%
Madsen%
\ \BBA {} Winkler%
}{%
{\protect \APACyear {1982}}%
}]{%
madsen1982intraductal}
\APACinsertmetastar {%
madsen1982intraductal}%
\begin{APACrefauthors}%
Madsen, P.%
\BCBT {}\ \BBA {} Winkler, K.%
\end{APACrefauthors}%
\unskip\
\newblock
\APACrefYearMonthDay{1982}{}{}.
\newblock
{\BBOQ}\APACrefatitle {The intraductal pancreatic pressure in chronic
  obstructive pancreatitis} {The intraductal pancreatic pressure in chronic
  obstructive pancreatitis}.{\BBCQ}
\newblock
\APACjournalVolNumPages{Scandinavian Journal of
  Gastroenterology}{17}{4}{553--554,}
\newblock

\newblock

\PrintBackRefs{\CurrentBib}

\bibitem [\protect \citeauthoryear {%
Mates%
, Gupta%
, Bell%
\BCBL {}\ \BBA {} Klocke%
}{%
Mates%
\ \protect \BOthers {.}}{%
{\protect \APACyear {1978}}%
}]{%
mates1978fluid}
\APACinsertmetastar {%
mates1978fluid}%
\begin{APACrefauthors}%
Mates, R.E.%
, Gupta, R.L.%
, Bell, A.C.%
\BCBL {} Klocke, F.J.%
\end{APACrefauthors}%
\unskip\
\newblock
\APACrefYearMonthDay{1978}{}{}.
\newblock
{\BBOQ}\APACrefatitle {Fluid dynamics of coronary artery stenosis.} {Fluid
  dynamics of coronary artery stenosis.}{\BBCQ}
\newblock
\APACjournalVolNumPages{Circulation research}{42}{1}{152--162,}
\newblock

\newblock

\PrintBackRefs{\CurrentBib}

\bibitem [\protect \citeauthoryear {%
Matos%
\ \protect \BOthers {.}}{%
Matos%
\ \protect \BOthers {.}}{%
{\protect \APACyear {1997}}%
}]{%
matos1997pancreatic}
\APACinsertmetastar {%
matos1997pancreatic}%
\begin{APACrefauthors}%
Matos, C.%
, Metens, T.%
, Devi{\`e}re, J.%
, Nicaise, N.%
, Braude, P.%
, Van~Yperen, G.%
\BDBL {}Struyven, J.%
\end{APACrefauthors}%
\unskip\
\newblock
\APACrefYearMonthDay{1997}{}{}.
\newblock
{\BBOQ}\APACrefatitle {Pancreatic duct: morphologic and functional evaluation
  with dynamic MR pancreatography after secretin stimulation.} {Pancreatic
  duct: morphologic and functional evaluation with dynamic mr pancreatography
  after secretin stimulation.}{\BBCQ}
\newblock
\APACjournalVolNumPages{Radiology}{203}{2}{435--441,}
\newblock

\newblock

\PrintBackRefs{\CurrentBib}

\bibitem [\protect \citeauthoryear {%
Meyer%
\ \protect \BOthers {.}}{%
Meyer%
\ \protect \BOthers {.}}{%
{\protect \APACyear {2017}}%
}]{%
meyer2017predictive}
\APACinsertmetastar {%
meyer2017predictive}%
\begin{APACrefauthors}%
Meyer, K.%
, Ostrenko, O.%
, Bourantas, G.%
, Morales-Navarrete, H.%
, Porat-Shliom, N.%
, Segovia-Miranda, F.%
\BDBL {}others%
\end{APACrefauthors}%
\unskip\
\newblock
\APACrefYearMonthDay{2017}{}{}.
\newblock
{\BBOQ}\APACrefatitle {A predictive 3D multi-scale model of biliary fluid
  dynamics in the liver lobule} {A predictive 3d multi-scale model of biliary
  fluid dynamics in the liver lobule}.{\BBCQ}
\newblock
\APACjournalVolNumPages{Cell systems}{4}{3}{277--290,}
\newblock

\newblock

\PrintBackRefs{\CurrentBib}

\bibitem [\protect \citeauthoryear {%
Mittal%
\ \protect \BOthers {.}}{%
Mittal%
\ \protect \BOthers {.}}{%
{\protect \APACyear {2016}}%
}]{%
mittal2016computational}
\APACinsertmetastar {%
mittal2016computational}%
\begin{APACrefauthors}%
Mittal, R.%
, Seo, J.H.%
, Vedula, V.%
, Choi, Y.J.%
, Liu, H.%
, Huang, H.H.%
\BDBL {}George, R.T.%
\end{APACrefauthors}%
\unskip\
\newblock
\APACrefYearMonthDay{2016}{}{}.
\newblock
{\BBOQ}\APACrefatitle {Computational modeling of cardiac hemodynamics: Current
  status and future outlook} {Computational modeling of cardiac hemodynamics:
  Current status and future outlook}.{\BBCQ}
\newblock
\APACjournalVolNumPages{Journal of Computational Physics}{305}{}{1065--1082,}
\newblock

\newblock

\PrintBackRefs{\CurrentBib}

\bibitem [\protect \citeauthoryear {%
Modi%
\ \protect \BOthers {.}}{%
Modi%
\ \protect \BOthers {.}}{%
{\protect \APACyear {2018}}%
}]{%
modi2018optimal}
\APACinsertmetastar {%
modi2018optimal}%
\begin{APACrefauthors}%
Modi, B.N.%
, Ryan, M.%
, Chattersingh, A.%
, Eruslanova, K.%
, Ellis, H.%
, Gaddum, N.%
\BDBL {}Perera, D.%
\end{APACrefauthors}%
\unskip\
\newblock
\APACrefYearMonthDay{2018}{}{}.
\newblock
{\BBOQ}\APACrefatitle {Optimal Application of Fractional Flow Reserve to Assess
  Serial Coronary Artery Disease: A 3D-Printed Experimental Study With Clinical
  Validation} {Optimal application of fractional flow reserve to assess serial
  coronary artery disease: A 3d-printed experimental study with clinical
  validation}.{\BBCQ}
\newblock
\APACjournalVolNumPages{Journal of the American Heart
  Association}{7}{20}{e010279,}
\newblock

\newblock

\PrintBackRefs{\CurrentBib}

\bibitem [\protect \citeauthoryear {%
Moreau%
, Patel%
, Rosenkranz%
\BCBL {}\ \BBA {} Feng%
}{%
Moreau%
\ \protect \BOthers {.}}{%
{\protect \APACyear {2015}}%
}]{%
ComputationalModelingofPancreaticDuctFluidFlow_previous}
\APACinsertmetastar {%
ComputationalModelingofPancreaticDuctFluidFlow_previous}%
\begin{APACrefauthors}%
Moreau, C.%
, Patel, S.%
, Rosenkranz, L.%
\BCBL {} Feng, Y.%
\end{APACrefauthors}%
\unskip\
\newblock
\APACrefYearMonthDay{2015}{07}{}.
\newblock
\APACrefbtitle {Computational Modeling of Pancreatic Duct Fluid Flow.}
  {Computational modeling of pancreatic duct fluid flow.}
\PrintBackRefs{\CurrentBib}

\bibitem [\protect \citeauthoryear {%
Nolte%
\ \protect \BOthers {.}}{%
Nolte%
\ \protect \BOthers {.}}{%
{\protect \APACyear {2021}}%
}]{%
nolte2021validation}
\APACinsertmetastar {%
nolte2021validation}%
\begin{APACrefauthors}%
Nolte, D.%
, Urbina, J.%
, Sotelo, J.%
, Sok, L.%
, Montalba, C.%
, Valverde, I.%
\BDBL {}Bertoglio, C.%
\end{APACrefauthors}%
\unskip\
\newblock
\APACrefYearMonthDay{2021}{}{}.
\newblock
{\BBOQ}\APACrefatitle {Validation of 4D flow based relative pressure maps in
  aortic flows} {Validation of 4d flow based relative pressure maps in aortic
  flows}.{\BBCQ}
\newblock
\APACjournalVolNumPages{Medical image analysis}{74}{}{102195,}
\newblock

\newblock

\PrintBackRefs{\CurrentBib}

\bibitem [\protect \citeauthoryear {%
Pandol%
}{%
Pandol%
}{%
{\protect \APACyear {2010}}%
}]{%
pandol2010water}
\APACinsertmetastar {%
pandol2010water}%
\begin{APACrefauthors}%
Pandol, S.%
\end{APACrefauthors}%
\unskip\
\newblock
\APACrefYearMonthDay{2010}{}{}.
\newblock
{\BBOQ}\APACrefatitle {Water and ion secretion from the pancreatic ductal
  system} {Water and ion secretion from the pancreatic ductal system}.{\BBCQ}
\newblock
\APACjournalVolNumPages{The Exocrine Pancreas. Morgan \& Claypool Life
  Sciences}{}{}{,}
\newblock

\newblock

\PrintBackRefs{\CurrentBib}

\bibitem [\protect \citeauthoryear {%
Peng%
\ \protect \BOthers {.}}{%
Peng%
\ \protect \BOthers {.}}{%
{\protect \APACyear {2024}}%
}]{%
peng2024analysis}
\APACinsertmetastar {%
peng2024analysis}%
\begin{APACrefauthors}%
Peng, T.%
, Zhong, Y.%
, Lin, X.%
, Jiang, B.%
, Wang, P.%
\BCBL {} Jia, Y.%
\end{APACrefauthors}%
\unskip\
\newblock
\APACrefYearMonthDay{2024}{}{}.
\newblock
{\BBOQ}\APACrefatitle {Analysis and numerical investigation of bile flow
  dynamics within the strictured biliary duct} {Analysis and numerical
  investigation of bile flow dynamics within the strictured biliary
  duct}.{\BBCQ}
\newblock
\APACjournalVolNumPages{International Journal for Numerical Methods in
  Biomedical Engineering}{40}{2}{e3790,}
\newblock

\newblock

\PrintBackRefs{\CurrentBib}

\bibitem [\protect \citeauthoryear {%
Qiao%
\ \protect \BOthers {.}}{%
Qiao%
\ \protect \BOthers {.}}{%
{\protect \APACyear {2019}}%
}]{%
qiao2019numerical}
\APACinsertmetastar {%
qiao2019numerical}%
\begin{APACrefauthors}%
Qiao, Y.%
, Zeng, Y.%
, Ding, Y.%
, Fan, J.%
, Luo, K.%
\BCBL {} Zhu, T.%
\end{APACrefauthors}%
\unskip\
\newblock
\APACrefYearMonthDay{2019}{}{}.
\newblock
{\BBOQ}\APACrefatitle {Numerical simulation of two-phase non-Newtonian blood
  flow with fluid-structure interaction in aortic dissection} {Numerical
  simulation of two-phase non-newtonian blood flow with fluid-structure
  interaction in aortic dissection}.{\BBCQ}
\newblock
\APACjournalVolNumPages{Computer methods in biomechanics and biomedical
  engineering}{22}{6}{620--630,}
\newblock

\newblock

\PrintBackRefs{\CurrentBib}

\bibitem [\protect \citeauthoryear {%
Qiu%
\ \protect \BOthers {.}}{%
Qiu%
\ \protect \BOthers {.}}{%
{\protect \APACyear {2025}}%
}]{%
qiu2025numerical}
\APACinsertmetastar {%
qiu2025numerical}%
\begin{APACrefauthors}%
Qiu, Y.%
, Tai, Y.%
, Li, Y.%
, Wei, Q.%
, Wu, H.%
\BCBL {} Li, K.%
\end{APACrefauthors}%
\unskip\
\newblock
\APACrefYearMonthDay{2025}{}{}.
\newblock
{\BBOQ}\APACrefatitle {Numerical assessment of portal pressure gradient (PPG)
  based on clinically measured hepatic venous pressure gradient (HVPG) for
  liver cirrhosis patients} {Numerical assessment of portal pressure gradient
  (ppg) based on clinically measured hepatic venous pressure gradient (hvpg)
  for liver cirrhosis patients}.{\BBCQ}
\newblock
\APACjournalVolNumPages{Journal of Biomechanics}{180}{}{112498,}
\newblock

\newblock

\PrintBackRefs{\CurrentBib}

\bibitem [\protect \citeauthoryear {%
Sato%
, Miyashita%
, Yamauchi%
\BCBL {}\ \BBA {} Matsuno%
}{%
Sato%
\ \protect \BOthers {.}}{%
{\protect \APACyear {1986}}%
}]{%
sato1986role}
\APACinsertmetastar {%
sato1986role}%
\begin{APACrefauthors}%
Sato, T.%
, Miyashita, E.%
, Yamauchi, H.%
\BCBL {} Matsuno, S.%
\end{APACrefauthors}%
\unskip\
\newblock
\APACrefYearMonthDay{1986}{}{}.
\newblock
{\BBOQ}\APACrefatitle {The role of surgical treatment for chronic pancreatitis}
  {The role of surgical treatment for chronic pancreatitis}.{\BBCQ}
\newblock
\APACjournalVolNumPages{Annals of surgery}{203}{3}{266,}
\newblock

\newblock

\PrintBackRefs{\CurrentBib}

\bibitem [\protect \citeauthoryear {%
Seo%
\ \protect \BOthers {.}}{%
Seo%
\ \protect \BOthers {.}}{%
{\protect \APACyear {2014}}%
}]{%
seo2014effect}
\APACinsertmetastar {%
seo2014effect}%
\begin{APACrefauthors}%
Seo, J.H.%
, Vedula, V.%
, Abraham, T.%
, Lardo, A.C.%
, Dawoud, F.%
, Luo, H.%
\BCBL {} Mittal, R.%
\end{APACrefauthors}%
\unskip\
\newblock
\APACrefYearMonthDay{2014}{}{}.
\newblock
{\BBOQ}\APACrefatitle {Effect of the mitral valve on diastolic flow patterns}
  {Effect of the mitral valve on diastolic flow patterns}.{\BBCQ}
\newblock
\APACjournalVolNumPages{Physics of fluids}{26}{12}{,}
\newblock

\newblock

\PrintBackRefs{\CurrentBib}

\bibitem [\protect \citeauthoryear {%
Singh%
\ \protect \BOthers {.}}{%
Singh%
\ \protect \BOthers {.}}{%
{\protect \APACyear {2023}}%
}]{%
singh2023pancreatic}
\APACinsertmetastar {%
singh2023pancreatic}%
\begin{APACrefauthors}%
Singh, A.%
, Bush, N.%
, Bhullar, F.A.%
, Faghih, M.%
, Moreau, C.%
, Mittal, R.%
\BDBL {}others%
\end{APACrefauthors}%
\unskip\
\newblock
\APACrefYearMonthDay{2023}{}{}.
\newblock
{\BBOQ}\APACrefatitle {Pancreatic duct pressure: A review of technical aspects
  and clinical significance} {Pancreatic duct pressure: A review of technical
  aspects and clinical significance}.{\BBCQ}
\newblock
\APACjournalVolNumPages{Pancreatology}{23}{7}{858--867,}
\newblock

\newblock

\PrintBackRefs{\CurrentBib}

\bibitem [\protect \citeauthoryear {%
Smith%
, Pullan%
\BCBL {}\ \BBA {} Hunter%
}{%
Smith%
\ \protect \BOthers {.}}{%
{\protect \APACyear {2002}}%
}]{%
smith2002anatomically}
\APACinsertmetastar {%
smith2002anatomically}%
\begin{APACrefauthors}%
Smith, N.%
, Pullan, A.%
\BCBL {} Hunter, P.J.%
\end{APACrefauthors}%
\unskip\
\newblock
\APACrefYearMonthDay{2002}{}{}.
\newblock
{\BBOQ}\APACrefatitle {An anatomically based model of transient coronary blood
  flow in the heart} {An anatomically based model of transient coronary blood
  flow in the heart}.{\BBCQ}
\newblock
\APACjournalVolNumPages{SIAM Journal on Applied mathematics}{62}{3}{990--1018,}
\newblock

\newblock

\PrintBackRefs{\CurrentBib}

\bibitem [\protect \citeauthoryear {%
Tajikawa%
, Aoki%
\BCBL {}\ \BBA {} Fukuzawa%
}{%
Tajikawa%
\ \protect \BOthers {.}}{%
{\protect \APACyear {2023}}%
}]{%
tajikawa2023investigation}
\APACinsertmetastar {%
tajikawa2023investigation}%
\begin{APACrefauthors}%
Tajikawa, T.%
, Aoki, K.%
\BCBL {} Fukuzawa, H.%
\end{APACrefauthors}%
\unskip\
\newblock
\APACrefYearMonthDay{2023}{}{}.
\newblock
{\BBOQ}\APACrefatitle {Investigation of pancreatic juice reflux mechanism in
  high confluence of pancreaticobiliary ducts and pancreaticobiliary
  maljunction (Development and validation of a mathematical model for
  pancreatic and bile juice flow based on fluid mechanics)} {Investigation of
  pancreatic juice reflux mechanism in high confluence of pancreaticobiliary
  ducts and pancreaticobiliary maljunction (development and validation of a
  mathematical model for pancreatic and bile juice flow based on fluid
  mechanics)}.{\BBCQ}
\newblock
\APACjournalVolNumPages{Journal of Biorheology}{37}{2}{44--55,}
\newblock

\newblock

\PrintBackRefs{\CurrentBib}

\bibitem [\protect \citeauthoryear {%
Takayanagi%
\ \protect \BOthers {.}}{%
Takayanagi%
\ \protect \BOthers {.}}{%
{\protect \APACyear {2023}}%
}]{%
takayanagi2023clinical}
\APACinsertmetastar {%
takayanagi2023clinical}%
\begin{APACrefauthors}%
Takayanagi, T.%
, Sekino, Y.%
, Kasuga, N.%
, Ishii, K.%
, Nagase, H.%
\BCBL {} Nakajima, A.%
\end{APACrefauthors}%
\unskip\
\newblock
\APACrefYearMonthDay{2023}{}{}.
\newblock
{\BBOQ}\APACrefatitle {Clinical features and prognostic impact of pancreatic
  ductal adenocarcinoma without dilatation of the main pancreatic duct: A
  single-center retrospective analysis} {Clinical features and prognostic
  impact of pancreatic ductal adenocarcinoma without dilatation of the main
  pancreatic duct: A single-center retrospective analysis}.{\BBCQ}
\newblock
\APACjournalVolNumPages{Diagnostics}{13}{5}{963,}
\newblock

\newblock

\PrintBackRefs{\CurrentBib}

\bibitem [\protect \citeauthoryear {%
Taylor%
, Fonte%
\BCBL {}\ \BBA {} Min%
}{%
Taylor%
\ \protect \BOthers {.}}{%
{\protect \APACyear {2013}}%
}]{%
taylor2013computational}
\APACinsertmetastar {%
taylor2013computational}%
\begin{APACrefauthors}%
Taylor, C.A.%
, Fonte, T.A.%
\BCBL {} Min, J.K.%
\end{APACrefauthors}%
\unskip\
\newblock
\APACrefYearMonthDay{2013}{}{}.
\newblock
{\BBOQ}\APACrefatitle {Computational fluid dynamics applied to cardiac computed
  tomography for noninvasive quantification of fractional flow reserve:
  scientific basis} {Computational fluid dynamics applied to cardiac computed
  tomography for noninvasive quantification of fractional flow reserve:
  scientific basis}.{\BBCQ}
\newblock
\APACjournalVolNumPages{Journal of the American College of
  Cardiology}{61}{22}{2233--2241,}
\newblock

\newblock

\PrintBackRefs{\CurrentBib}

\bibitem [\protect \citeauthoryear {%
Vandenbulcke%
, De~Pauw%
, Dewaele%
, Degroote%
\BCBL {}\ \BBA {} Segers%
}{%
Vandenbulcke%
\ \protect \BOthers {.}}{%
{\protect \APACyear {2022}}%
}]{%
vandenbulcke2022computational}
\APACinsertmetastar {%
vandenbulcke2022computational}%
\begin{APACrefauthors}%
Vandenbulcke, S.%
, De~Pauw, T.%
, Dewaele, F.%
, Degroote, J.%
\BCBL {} Segers, P.%
\end{APACrefauthors}%
\unskip\
\newblock
\APACrefYearMonthDay{2022}{}{}.
\newblock
{\BBOQ}\APACrefatitle {Computational fluid dynamics model to predict the
  dynamical behavior of the cerebrospinal fluid through implementation of
  physiological boundary conditions} {Computational fluid dynamics model to
  predict the dynamical behavior of the cerebrospinal fluid through
  implementation of physiological boundary conditions}.{\BBCQ}
\newblock
\APACjournalVolNumPages{Frontiers in bioengineering and
  biotechnology}{10}{}{1040517,}
\newblock

\newblock

\PrintBackRefs{\CurrentBib}

\bibitem [\protect \citeauthoryear {%
Vondrasek%
\ \BBA {} Eberhardt%
}{%
Vondrasek%
\ \BBA {} Eberhardt%
}{%
{\protect \APACyear {1974}}%
}]{%
vondrasek1974semiconductors}
\APACinsertmetastar {%
vondrasek1974semiconductors}%
\begin{APACrefauthors}%
Vondrasek, P.%
\BCBT {}\ \BBA {} Eberhardt, G.%
\end{APACrefauthors}%
\unskip\
\newblock
\APACrefYearMonthDay{1974}{}{}.
\newblock
{\BBOQ}\APACrefatitle {Semiconductors in the recording of pressure during
  endoscopy. Preliminary report on the technic of measurement} {Semiconductors
  in the recording of pressure during endoscopy. preliminary report on the
  technic of measurement}.{\BBCQ}
\newblock
\APACjournalVolNumPages{Zeitschrift fur Gastroenterologie}{12}{6}{453--458,}
\newblock

\newblock

\PrintBackRefs{\CurrentBib}

\bibitem [\protect \citeauthoryear {%
Yamashita%
, Ashida%
\BCBL {}\ \BBA {} Kitano%
}{%
Yamashita%
\ \protect \BOthers {.}}{%
{\protect \APACyear {2022}}%
}]{%
yamashita2022imaging}
\APACinsertmetastar {%
yamashita2022imaging}%
\begin{APACrefauthors}%
Yamashita, Y.%
, Ashida, R.%
\BCBL {} Kitano, M.%
\end{APACrefauthors}%
\unskip\
\newblock
\APACrefYearMonthDay{2022}{}{}.
\newblock
{\BBOQ}\APACrefatitle {Imaging of fibrosis in chronic pancreatitis} {Imaging of
  fibrosis in chronic pancreatitis}.{\BBCQ}
\newblock
\APACjournalVolNumPages{Frontiers in Physiology}{12}{}{800516,}
\newblock

\newblock

\PrintBackRefs{\CurrentBib}

\bibitem [\protect \citeauthoryear {%
Zhang%
\ \protect \BOthers {.}}{%
Zhang%
\ \protect \BOthers {.}}{%
{\protect \APACyear {2020}}%
}]{%
zhang2020deep}
\APACinsertmetastar {%
zhang2020deep}%
\begin{APACrefauthors}%
Zhang, J.%
, Zhu, L.%
, Yao, L.%
, Ding, X.%
, Chen, D.%
, Wu, H.%
\BDBL {}others%
\end{APACrefauthors}%
\unskip\
\newblock
\APACrefYearMonthDay{2020}{}{}.
\newblock
{\BBOQ}\APACrefatitle {Deep learning--based pancreas segmentation and station
  recognition system in EUS: Development and validation of a useful training
  tool (with video)} {Deep learning--based pancreas segmentation and station
  recognition system in eus: Development and validation of a useful training
  tool (with video)}.{\BBCQ}
\newblock
\APACjournalVolNumPages{Gastrointestinal endoscopy}{92}{4}{874--885,}
\newblock

\newblock

\PrintBackRefs{\CurrentBib}

\end{thebibliography}


\end{document}